\documentclass[useAMS,usenatbib]{mnras}
  \usepackage{amsmath}
\usepackage{amssymb}
\usepackage{graphicx}
\usepackage{blkarray} 

\DeclareMathOperator\arctanh{arctanh}

\title[Mesh-free free-form lensing I]
  {Mesh-free free-form lensing I: Methodology and application to mass reconstruction}
  
  \author[J.~Merten]{
  Julian~Merten $^{1,2,3}$\thanks{\href{julian.merten@physics.ox.ac.uk}{julian.merten@physics.ox.ac.uk}}
  \newauthor \\
  $^1$Department of Physics, University of Oxford, Keble Road, Oxford OX1 3RH, U.K.\\
  $^2$Jet Propulsion Laboratory, California Institute of Technology, 4800 Oak Grove Drive, Pasadena, CA 91109, USA\\
  $^3$California Institute of Technology, MC 249-17, Pasadena, CA 91125, USA
  }
  
  \date{\textcopyright~2015 California Institute of Technology. Government sponsorship acknowledged. All rights reserved.\\
  Submitted to the Monthly Notices of the Astronomical Society.}

\begin{document}
  \label{firstpage}
   \maketitle
   \begin{abstract}
    Many applications and algorithms in the field of gravitational lensing make use of meshes with a finite number of nodes
    to analyze and manipulate data. Specific examples in lensing are astronomical CCD images in general,
    the reconstruction of density distributions from lensing data, lens--source plane mapping or the characterization and
    interpolation of a point-spread-function.    
    We present a numerical framework to interpolate and differentiate in the mesh-free domain, defined by nodes
    with coordinates that follow no regular pattern. The framework is based on radial basis functions (RBFs) 
    to smoothly represent data around the nodes.
    We demonstrate the performance of Gaussian RBF-based, mesh-free interpolation and differentiation, 
    which reaches the sub-percent level in both cases. 
    We use our newly developed framework to translate ideas of free-form mass reconstruction from lensing onto the mesh-free domain. 
    By reconstructing a simulated mock lens 
    we find that strong lensing only reconstructions achieve $<10\%$ accuracy 
    in the areas where these constraints are available but provide poorer results when departing from these regions. 
    Weak-lensing only reconstructions give $<10\%$ accuracy outside the strong lensing regime,  
    but cannot resolve the inner core structure of the lens. Once both regimes are combined, 
    accurate reconstructions can be achieved over the full field of view. 
    The reconstruction of a simulated lens, using constraints that mimics real observations, yields accurate results in terms of surface-mass density, NFW parameter, Einstein radius and magnification map recovery,
    encouraging the application of this method to real data.
   \end{abstract}
   
   
   \section{Introduction}
   Many techniques of astrophysical data analysis, although they in principle work on smooth and continuous data, are
   confined to a discrete numerical domain, which evaluates the input data using a finite number of analysis nodes. This 
   numerical domain is usually referred to as a mesh and the coordinates of its nodes can have different dimensionality,
   depending on the application. 
   In many cases the structure of these node coordinates
   follows a regular pattern, so a constant separation of nodes in each dimension. This regular pattern is convenient because
   it reduces the numerical complexity of the problem and simplifies many numerical algorithms since the spatial structure of the
   data is highly symmetric and easy to implement. However, many applications need a more sophisticated description of the spatial 
   distribution of input data. A more general mesh layout
   is provided by adaptive mesh refinement (AMR) which increases the resolution of the mesh wherever this is allowed by the quality of the data and affordable in terms 
   of CPU-time. The distances between nodes in each dimension is now adaptive but still follows a regular pattern. 
   However, real astrophysical data, e.g. the 
   distribution of galaxies or stars in a certain patch of the sky can be distributed in a very irregular fashion with density fluctuations, clusters
   and voids. When translating the input data onto a regular, structured mesh, this can lead to highly oversampled and partly unconstrained meshes 
   or interpolation is needed, introducing the associated interpolation errors. 
   The other extreme are undersampled meshes where averaging techniques compress the data onto
   a regular mesh. This comes at the price of smoothing out information and thus not making use of the full potential of the data.
   
   In this work we present a framework which can deal with a mesh-free structure in the input data distribution. This means that the coordinates 
   of the nodes follow no regular pattern and can have any values within the numerical domain. Two important types of data manipulation on such structures
   are interpolation and differentiation and we will introduce efficient algorithms which can achieve both of these tasks
   in any spatial dimension. The key to such a framework are radial basis functions (RBFs), functions which only depend on the
   distance of their evaluation points to certain reference points.
   
   This work, which focuses on mass reconstruction from gravitational lensing, is only the first in a series, which will exploit our mesh-free numerical techniques.
   Two regimes are typically distinguished in lensing mass reconstruction.
   Strong lensing is usually confined to the inner-most core of the gravitational lens and produces
   spectacular observational constraints such as multiple images of the same source, gravitational arcs or even rings. The domain of
   weak lensing is further away from the center of the lens but spans large areas and manifests itself by the weak distortion in the shape of background
   galaxies behind the lens. Reconstruction techniques are divided into two classes, 
   although this distinction is by no means unique or even consistent in some cases. Parametric techniques \citep[e.g.][for some recent examples]{Kneib1996,Broadhurst2005,Smith2005,Halkola2006,Jullo2007,Zitrin2009,Oguri2010,Newman2013,Jullo2014,Monna2014,Johnson2014a} assume
   a parametric form of the underlying mass density distribution for the lens and typically make the assumption that light traces mass
   in the positioning of these parametric forms. On the other hand, free-form\footnote{Sometimes also dubbed as nonparametric, which is misleading.
   In fact every mesh node is usually a free parameter rendering these methods highly parametric.} methods \citep[see e.g.][for some recent examples]{Broadhurst1995,Bartelmann1996,Abdelsalam1998,Bridle1998,Seitz1998,Bradav2005,Cacciato2006,Liesenborgs2006,Diego2007,Jee2007,Coe2008,Bradav2009,Merten2009,Williams2011,Merten2011, Merten2015,Diego2015} usually do not make this assumption
   and purely rely on the input data either based on weak lensing, 
   strong lensing or a combination of the two. This is possible while using a reconstruction mesh and directly inverting the underlying
   equations describing lensing on this mesh. In the following, we introduce a free-form method combining weak and strong lensing, 
   which uses our new mesh-free numerical framework. This method translates original ideas by \citet{Bartelmann1996},
    \citet{Seitz1998}, \citet{Bradav2005}, \citet{Cacciato2006} and \citet{Merten2009} into the flexible and efficient mesh-free numerical domain.
    Alternative implementations of such ideas can e.g. be found in \citet{Bradav2009}.
    
    This work is structured as follows: In Sec.~\ref{sec_rbf} we introduce RBFs and how they can be used to numerically 
    interpolate and differentiate. In Sec.~\ref{sec_lensing} we use the developed techniques to implement 
    a free-form reconstruction algorithm that can be used in the mesh-free domain and consistently combines the regimes of weak and
    strong gravitational lensing. We test our implementation with numerically simulated data 
    in Sec.~\ref{sec_testing} and we conclude in Sec.~\ref{sec_conclusions}. In App.~\ref{app_rbf_inter} we provide more details
    on the performance of interpolation with RBFs to the interested reader, as we do in App.~\ref{app_rbf_fd} for
    mesh-free numerical differentiation. App.~\ref{app_lse} gives some missing but not crucial details on the 
    concrete implementation of the reconstruction algorithm. In many graphical illustrations in this paper we have to visualize
    mesh-free data. We do so by using the Voronoi tessellation of the evaluation points and by assigning a 
    function value to each Voronoi cell, which refers to the function value at the respective coordinate. 
    Throughout this work, we assume a flat $\Lambda$CDM cosmology with $h=0.7$, $\Omega_{\Lambda}=0.7$ and $\Omega_{\textrm{m}}=0.3$. In Sec.~\ref{sec_ray}, we refer to physical quantities of 
      a simulated lens at $z=0.25$ where one arc second corresponds to 3.91 kpc.

   \section{Radial basis functions}
   \label{sec_rbf}

   In this general methodology section we will deal with functions which are defined on a 
   finite number of evaluation points. Based on this set of points, we will interpolate functions 
   in their numerical domain and calculate their derivatives. 
   The general idea which enables us to do so is based on RBFs, where we expand
   the discretely defined functions into a set of radially dependent functions around the evaluation points. For a thorough discussion
   of the concept of radial basis functions see \citet{Fasshauer2007} and references therein.
   
   \subsection{Unstructured meshes and mesh-free data}
   \label{subsec_unstructured_grids}
   We define a mesh $\mathcal{M}$ as a finite collection of N support points $\vec{x}$
   \begin{equation}
    \mathcal{M} = \left[\vec{x}_{1},\vec{x}_{2},...,\vec{x}_{\mathrm{N}}\right].
   \end{equation}
   The dimensionality of the mesh is given by the number of coordinates $n$ needed to define each support point 
   $\vec{x} = \left(x_{1},x_{2},...,x_{\mathrm{d}}\right)$. In most cases we will focus on the 2D case with $\vec{x} = \left(x_{1},x_{2}\right)$,
   but the methodology presented in this section generalizes to any dimension.
   
   This definition of a mesh $\mathcal{M}$ is general in the sense that it includes cases with regularly distributed nodes or randomly distributed nodes.
   In this work we will show examples for both kinds of meshes, but generally focus on unstructured, meaning irregularly shaped mesh configurations.
   The range of coordinates depends on the numerical domain the mesh is defined on but for simplicity we will mostly restrict ourselves to coordinates $x_{i}\in\left[-1,...,1\right]$ with $i$ depending on the dimensionality of the problem.
   Unstructured meshes can be still restricted when it comes to the distribution of their nodes. In the field of finite elements, for example, specific
   elements need to be formed with restrictions on the aspect ratios of their edges. For our purposes we do not have such restrictions, which is why
   we distinguish the notion of unstructured meshes from our, more general, case where the distribution of node coordinates follows no restrictions and which we call mesh-free.

   \subsection{Interpolation with radial basis functions}
   \label{subsec_rbf_interpol}
   We analyze a function $f$ defined at $n$ nodes $\vec{x}_{1},...,\vec{x}_{n}$ and  denote the values of the function at these points $f(\vec{x_{1}})$,...,$f(\vec{x}_{n})$ with $f_{1}$,...,$f_{n}$. 
   We use RBFs $\phi(\vec{x}) = \phi(\lVert\vec{x}-\vec{x}_{0}\rVert)$ to interpolate $f$ to any position $\vec{x}$ in the numerical domain.  
   Throughout this work, radial distances are defined as the Euclidean norm $L_{2}$ with the short-hand notation $\lVert\vec{x}-\vec{x}_{0}\rVert = r$ for a given reference point $\vec{x}_{0}$. Typical choices for RBFs are listed in Tab.~\ref{tab_RBFs}, but in this work we will restrict
   ourselves to Gaussian RBFs. 
   In this case, the only free parameter, $\epsilon$, is called the shape parameter and has to be chosen carefully as will be discussed in great detail in Sec.~\ref{subsec_shapeparam} and App.~\ref{app_rbf_inter}. 
   For a more thorough discussion of the underlying mathematical concepts we refer to \citet{Fornberg2011}, \citet{Larsson2013}, \citet{Fornberg2013627}, \citet{Flyer2014} and references therein. 
   
   \begin{table}
     \caption{A few common choices for radial basis functions}
     \label{tab_RBFs}
     \centering
     \begin{tabular}{@{}lc}
     \hline
      Name & functional form $\phi(r)$\\
      \hline
      Gaussian & $e^{-(\epsilon r)^2}$\\
      Multiquadric & $\sqrt{1+(\epsilon r)^{2}}$\\
      Inverse Multiquadric & $1/\sqrt{1+(\epsilon r)^{2}}$\\
      Inverse Quadric & $1/(1+(\epsilon r)^{2})$\\
      Polyharmonic Spline & \begin{tabular}{l} $r^{2m-1}$ \\$r^{2m}\log r$ \end{tabular} with $m\in \mathbb{N}$\\
     \hline     
     \end{tabular}
    \end{table}

\begin{figure}
      \includegraphics[width=.5\textwidth]{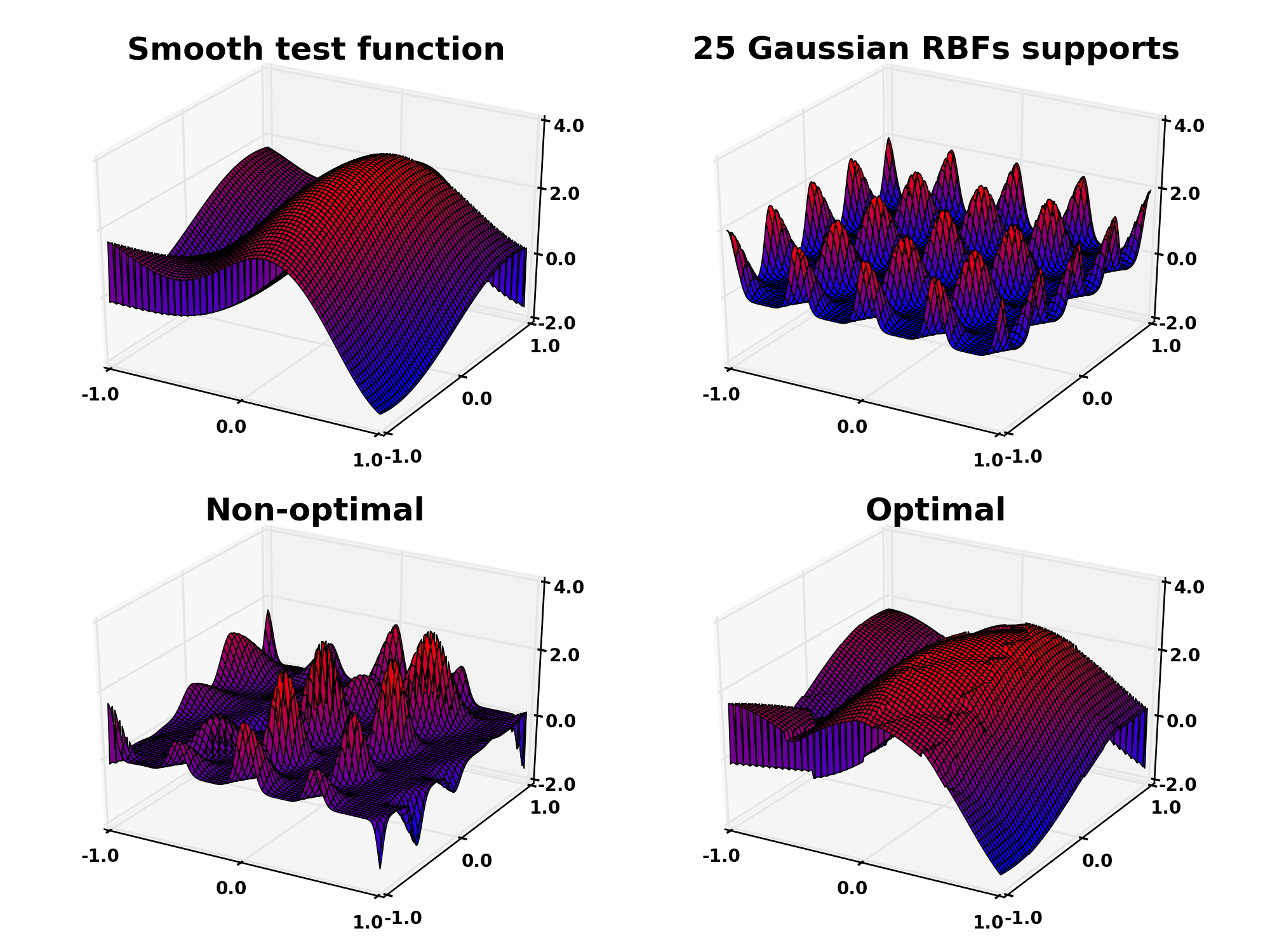}
     \caption{A graphical illustration of expanding a 2D function into a set of 25 Gaussian RBFs. The top left panel shows the function
     to be interpolated defined on a unit mesh. The top right panels shows 25 Gaussian RBFs with an arbitrary amplitude of 2, an arbitrary shape parameter of 10
     and placed on regularly spaced evaluation points. In the bottom left panel the amplitude of the Gaussians was set to the actual value
     of the test function at the position of the evaluation points, but the shape parameter was not optimized. In the bottom right panel, the amplitudes
     of the Gaussians were chosen by solving Eq.~\ref{equ_rbf_interpol_system} and the shape parameter was optimized for this interpolation problem.}
     \label{fig_rbf_gaussians}
    \end{figure}

   We write the interpolant $\tilde{f}$ of the function $f$ defined at nodes $\vec{x}_{1},...,\vec{x}_{n}$ as a weighted sum over RBFs 
   \begin{equation}
   \label{equ_rbf_interpol}
    \tilde{f}(\vec{x})=\sum\limits_{i=1}^{n}\lambda_{i}\phi(\lVert\vec{x}-\vec{x}_{i}\rVert)
   \end{equation}
   with weighting coefficients  $\lambda_{i}$. Because of $\tilde{f}(\vec{x}_{i})=f_{i}$ we can calculate the coefficients $\lambda$ by solving the linear system of equations
   \begin{equation}
   \label{equ_rbf_interpol_system}
   \underbrace{
    \begin{pmatrix}
     \phi(\lVert\vec{x}_{1}-\vec{x}_{1}\rVert)&\hdots&\phi(\lVert\vec{x}_{1}-\vec{x}_{n}\rVert)\\
     \vdots&\ddots&\vdots\\
     \phi(\lVert\vec{x}_{n}-\vec{x}_{1}\rVert)&\hdots&\phi(\lVert\vec{x}_{n}-\vec{x}_{n}\rVert)
    \end{pmatrix}}_{\mathcal{F}}
    \begin{pmatrix}
    \lambda_{1}\\
    \vdots\\
    \lambda_{n}
    \end{pmatrix}
    =
    \begin{pmatrix}
    f_{1}\\
    \vdots\\
    f_{n}
    \end{pmatrix}
   \end{equation}
    It is important to note that for distinct nodes, this linear system cannot become singular,
    no matter how the nodes are scattered in any number of dimensions.
    Fig.~\ref{fig_rbf_gaussians} shows a graphical illustration of the expansion of a function into Gaussian RBFs around evaluation points. The accuracy of the interpolation is 
    certainly dependent on the shape parameter and on the choice and number of nodes. We discuss the shape parameter further in Sec.~\ref{subsec_shapeparam} and present a detailed
    performance analysis of the RBF-based interpolation scheme in App.~\ref{app_rbf_inter}.
    
   \subsection{Mesh-free numerical derivatives}
   \label{subsec_rbf_fd}
   Numerical derivatives are usually calculated by means of finite differencing (FD)
   \begin{equation}
   \label{equ_FD}
    Df(\vec{x}) \approx \sum\limits_{i=1}^{n}w_{i}f(\vec{x}_{i})
   \end{equation}
   for a linear differential operator $D$ and with evaluation points $\vec{x}_{i}$ defining a FD stencil. 
   There are several ways of finding the weights $w$ for the FD on regular meshes, ranging from 
   the Lagrange interpolation polynomial, Taylor expansion, monomial test functions to the elegant Pad\'{e}-algorithm. For a thorough
   description of all different techniques and the original references, see \citet{Fornberg1998}. 
   Here we focus on monomial test functions since it will set
   the stage for mesh-free RBF-based FD later.
   
   The motivation for the use of monomial test functions is to enforce that Eq.~\ref{equ_FD} holds exactly when $f$ is a polynomial of degree
   n-1. In the 1D case, the weights in Eq.~\ref{equ_FD} are then given by the solution of the linear system of equations
   \begin{equation}
   \label{equ_mono_lse}
    \begin{pmatrix}
     1&1&\hdots&1\\
     x_{1}&x_{2}&\hdots&x_{n}\\
     \vdots&\vdots&\ddots&\vdots\\
     x^{n-1}_{1}&x^{n-1}_{2}&\hdots&x^{n-1}_{n}\\
    \end{pmatrix}
    \begin{pmatrix}
    w_{1}\\
    w_{2}\\
    \vdots\\
    w_{n}
    \end{pmatrix}
    =
    \begin{pmatrix}
    D1(\vec{x})\\
    Dx(\vec{x})\\
    \vdots\\
    Dx^{n-1}(\vec{x})
    \end{pmatrix}
   \end{equation}
   where $x_{1},...,x_{n}$ are the evaluation points of the finite-differencing stencil and the differential operator $D$ is applied
   to the monomial test functions $x^{n-1}$ at the point of interest $\vec{x}$. This
   approach easily generalizes to higher dimensions by inserting also the other coordinate components of $\vec{x}$ and 
   mixed-component monomial test functions.
   The case of $D=\bigtriangleup$ and $n=5$ recovers the well known finite-differencing stencil on a regular mesh
   with node separation $h$, evaluated at $\vec{x}=\vec{x}_{0}$

   \begin{equation}
   \label{equ_2d_laplace}
   h^{-2}
       \begin{pmatrix}
        1&1&-4&1&1
       \end{pmatrix}
	\begin{pmatrix}
	f\left(x_{0}+h,y_{0}\right)\\
	f\left(x_{0}-h,y_{0}\right)\\
	f(\vec{x}_{0})\\
	f\left(x_{0},y_{0}+h\right)\\
	f\left(x_{0},y_{0}-h\right)\
	\end{pmatrix}
\approx \bigtriangleup f(\vec{x}_{0}).
   \end{equation}
   
   Inspired by this approach, we substitute the monomial test functions with RBFs and advance to a mesh-free formulation by
   centering the RBFs on the evaluation points of the function. In complete analogy to Eq.~\ref{equ_mono_lse}, this Ansatz
   leads to the following linear system of equations to find the finite differencing weights
  \begin{equation}
  \label{equ_rbf_fd_lse}
\mathcal{F}
    \begin{pmatrix}
    w_{1}\\
    \vdots\\
    w_{n}
    \end{pmatrix}
    =
    \begin{pmatrix}
    D\phi(\lVert\vec{x}-\vec{x}_{1}\lVert)\\
    \vdots\\
    D\phi(\lVert\vec{x}-\vec{x}_{n}\lVert)
    \end{pmatrix}.   
  \end{equation}
  Monomial terms can be included again in order to increase the accuracy of the numerical derivatives. 
  In the 2D case with monomial terms up to second order the FD weights are given by the solution to 
    \begin{align}
  \label{equ_rbf_lse_mono}
  \begin{pmatrix}
   &1&x_{1}&y_{1}&x^{2}_{1}&y^{2}_{1}&x_{1}y_{1}\\
   \mathcal{F}&\vdots&\vdots&\vdots&\vdots&\vdots&\vdots\\
   ~&1&x_{n}&y_{n}&x^{2}_{n}&y^{2}_{n}&x_{n}y_{n}\\
   1\hdots1&&&&&&\\
   x_{1}\hdots x_{n}&&&&&&\\
   y_{1}\hdots y_{n}&&&&&&\\
   x^{2}_{1}\hdots x^{2}_{n}&&&0&&&\\
   y^{2}_{1}\hdots y^{2}_{n}&&&&&&\\
   x_{1}y_{1}\hdots y_{n}y_{n}&&&&&&\\
  \end{pmatrix}
  \begin{pmatrix}
    w_{1}\\
    \vdots\\
    w_{n}\\
    w_{n+1}\\
    w_{n+2}\\
    w_{n+3}\\
    w_{n+4}\\
    w_{n+5}\\
    w_{n+6}
    \end{pmatrix}
    = \nonumber \\ 
    \begin{pmatrix}
    D\phi(\lVert\vec{x}-\vec{x}_{1}\lVert)\\
    \vdots\\
    D\phi(\lVert\vec{x}-\vec{x}_{n}\lVert)\\
    D1(\vec{x})\\
    Dx(\vec{x})\\
    Dy(\vec{x})\\
    Dx^{2}(\vec{x})\\
    Dy^{2}(\vec{x})\\
    D(xy)(\vec{x})
    \end{pmatrix}   
  \end{align}
  Only $w_{1},...,w_{n}$ are used as weights in Eq.~\ref{equ_FD}. The remaining values $w_{n+1},...,w_{n+6}$ have no obvious meaning. 
  This 2D RBF FD scheme with up to second order monomial terms applies to most of our applications in this paper and is
  the scheme that we use in this work from now on, if not otherwise stated. 
  The performance of RBF-derived FD stencils is discussed in the next section and in great detail in App.~\ref{app_rbf_fd}. 
  
   \subsection{Discussion of the shape parameter}
   \label{subsec_shapeparam}
\begin{figure}
    \centering
      \includegraphics[width=.475\textwidth]{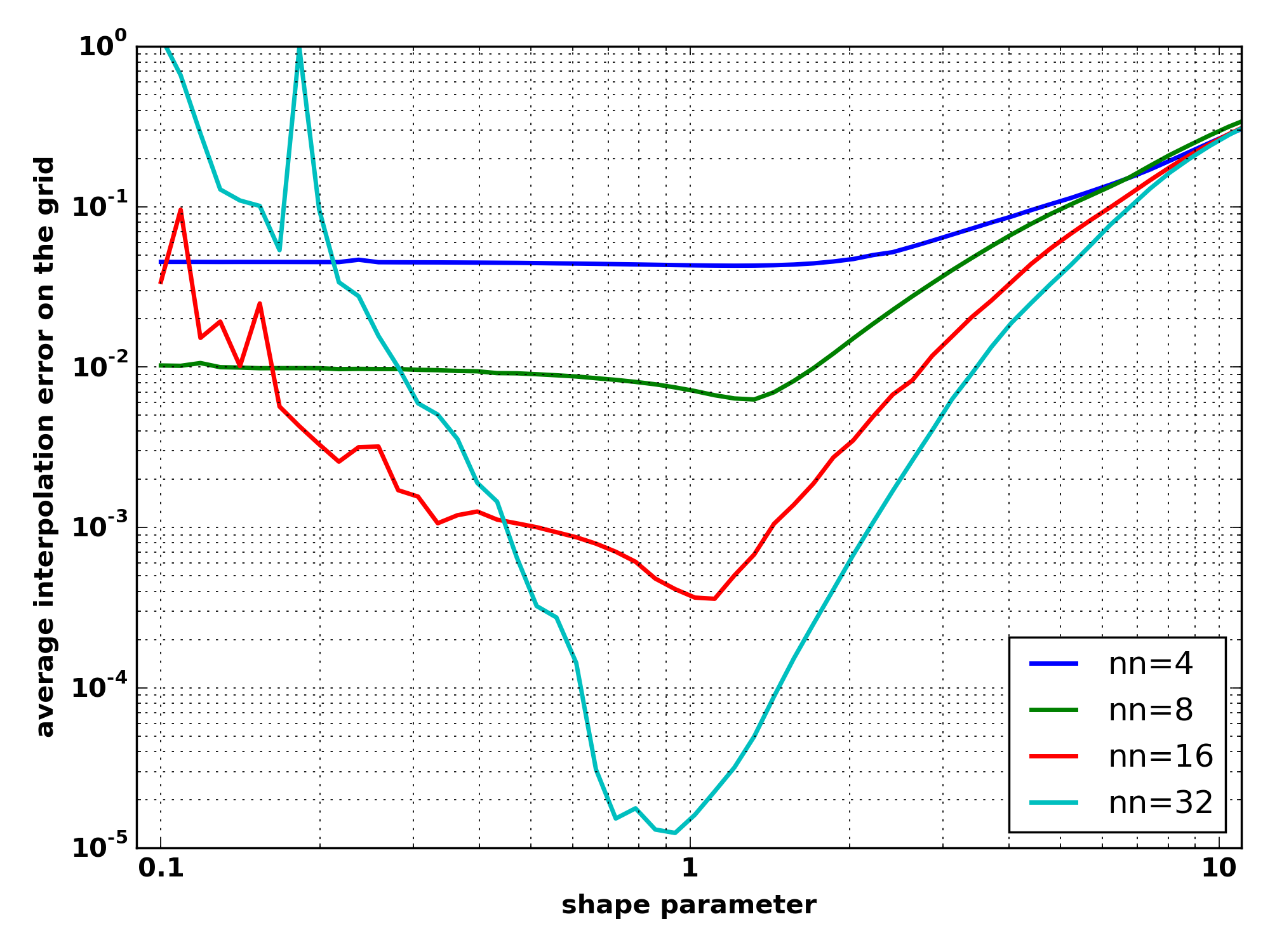}
     \caption{The accuracy of the test function interpolation as a function of the shape parameter and of the number of nearest neighbors 
     used to perform the interpolation. The interpolation mesh consists of 900 nodes and of 180 support points. The interpolation
     function is shown in Eq.~\ref{equ_test_function}.}
       \label{fig_shape_param_inter}
    \end{figure}

   The shape parameter controls the width of the Gaussian RBF and crucially controls 
   the accuracy of the RBF application no matter if one uses RBFs for interpolation or FD. For a complete 
   discussion of the shape parameter we refer the interested reader to the work by \citet{Fornberg2011}, \citet{Fornberg2013627} and \citet{Larsson2013}. 
   In the following we will heuristically analyze the effect of the shape parameter by introducing the test function
  \begin{equation}
   \label{equ_test_function}
   f(x,y)=1+\sin(4x)+\cos(3x)+\sin(2y).
  \end{equation}
  Based on 180 randomly chosen evaluation points, we interpolate the test 
  function to 900 random locations (nodes) on the unit disk. For a more complete description of the test setup we refer the interested
  reader to App.~\ref{app_rbf_inter}. 
  In Fig.~\ref{fig_shape_param_inter} we show the shape parameter dependence of the average and maximum relative error of the interpolation
  and also include the dependence on the number of nearest neighbor evaluation points used to carry out the interpolation
  for each interpolant evaluation point. 
  As one can clearly see, there is an optimal choice
  for the shape parameter in order to achieve maximum accuracy. This choice depends on the actual node coordinates, 
  and the number of evaluation points. Once the shape parameter is optimized \citep{Fornberg2013627}, very high accuracies in the interpolation
  can be achieved, reaching an average relative interpolation error $\sim 10^{-5}$.  We discuss this in more detail 
  in App.~\ref{app_rbf_inter}, with another test function that is specifically chosen for our purposes. As a final remark on the theory of RBF interpolation we point out, that the approach
  can be further optimized by choosing a spatially varying shape parameter.
  Since the choice of a variable shape parameter, depending on the distribution of evaluation points, is not trivial \citep{Fornberg2007}
  we focus for now on the special case of a single $\epsilon$ value. We will present a more general
  method, that adaptively varies the shape parameter, in the course of the development of our mesh-free methods. 
    
\begin{figure}
      \includegraphics[width=.475\textwidth]{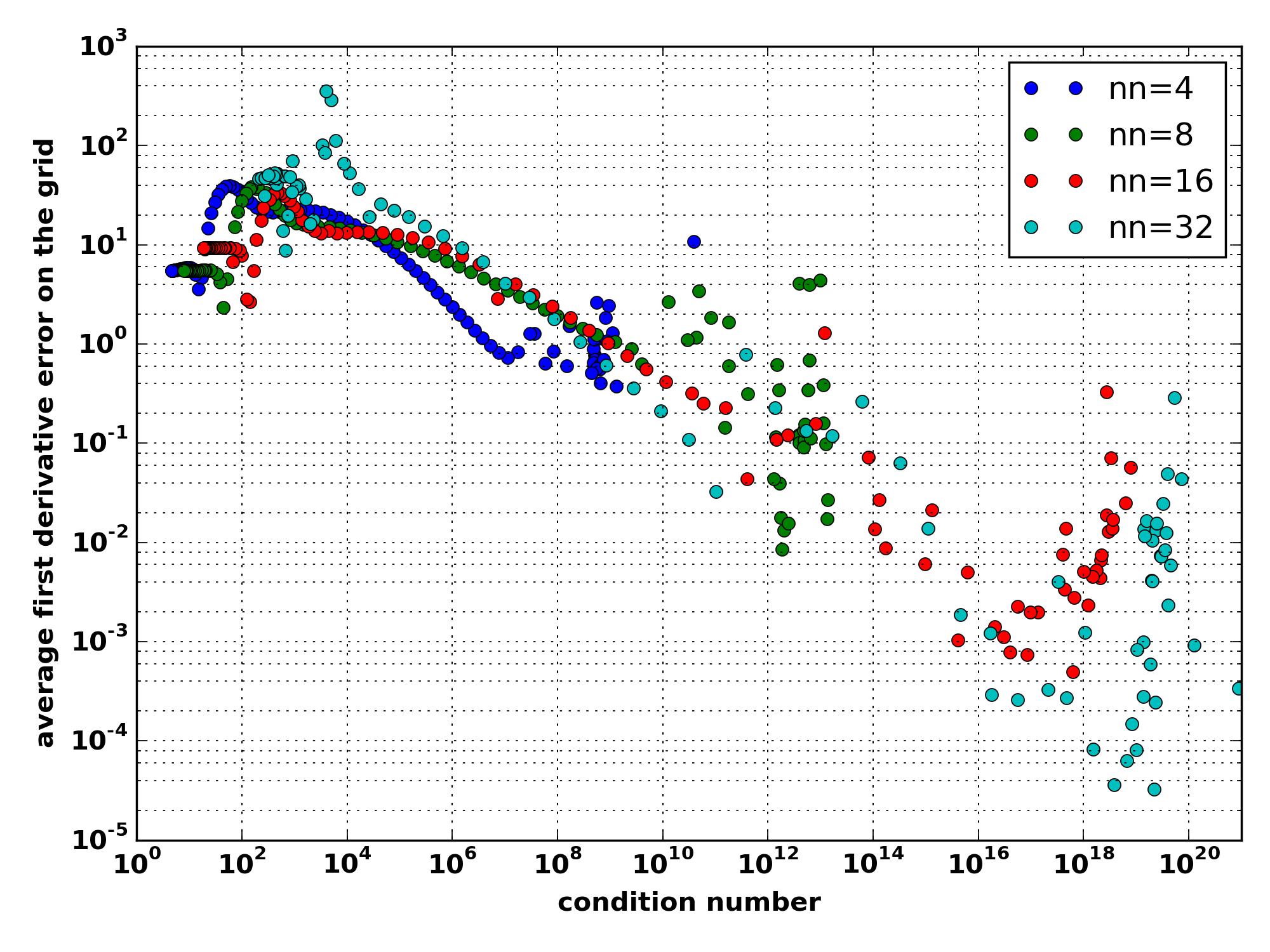}
     \caption{The accuracy of the first numerical $x$ derivative of the test function (Eq.~\ref{equ_test_function})
     as a function of the condition number and the number of nearest neighbors in the FD stencil.}
     \label{fig_shape_param_fd}
    \end{figure}

    For RBF-based FD stencils, it has been shown that the respective FD weights 
    only give accurate results if the condition number $C$\footnote{The condition number of the coefficient matrix is defined
    as the ratio between the largest and smallest singular value of the SVD decomposition of the coefficient matrix.} of the coefficient matrix in Eq.~\ref{equ_rbf_fd_lse} is close
    to critical\footnote{This means that $\log(C) \gtrsim\epsilon$, where $\epsilon$ is the machine precision of your numerical implementation.}, and thus depends on the machine precision of the implementation \citep{Fornberg2013627}. 
    One has to carefully monitor this condition number, which can be directly controlled by the shape parameter. 
    In Fig.~\ref{fig_shape_param_fd} we show the accuracy of the numerical first $x$ derivative 
    as a function of this condition number and the number of nearest neighbors in the stencil.
    The accuracy is rapidly increasing when approaching
    the ill-condition point, where $C$ is approaching the critical value, of the system. Once the system is ill-conditioned, or close to that, the accuracy is decreasing again and
    shows irregular behavior. This is discussed in much more detail in \citet{Fornberg2013627} and we give a more detailed heuristic assessment in App.~\ref{app_rbf_fd}. 
    It is also worth noting that the condition number is independent of the differential operator $D$, as can be seen from Eq.~\ref{equ_rbf_fd_lse}.
    In the following we mostly use 16 to 32 nearest neighbors to define the FD stencil in each evaluation point and therefore 
    we adjust the shape parameter to keep the condition number of $\mathcal{F}$ in the $10^{16}$--$10^{17}$ region.
    In order to avoid the dependence on shape parameter altogether, the use of polyharmonic spline (PHS) type RBFs (compare Tab.~\ref{tab_RBFs}) is a sensible approach.
    The performance of PHS-RBFs in RBF-FD applications is currently under investigation and shows promising results (Bengt Fornberg, private
    communication). We will implement these RBFs in the numerical framework that we present in the next section and will perform
    a series of tests to analyze the accuracy and feasibility of PHS-type RBFs for our purposes. 
    
\subsection{Numerical implementation}
\label{subsec_num_implementation}

   We briefly describe our own numerical implementation\footnote{\href{https://bitbucket.org/jmerten82/libmfree}{https://bitbucket.org/jmerten82/libmfree}} of mesh-free interpolation and differentiation with 
   RBFs.
 The library is written in C++ and mainly provides, among several helper routines, two classes. The central class 
 describes a collection of $n$ nodes with arbitrary coordinates in either 1D, 2D or 3D and is initially defined by an input vector 
 $X = \left[\vec{x}_{1}, \vec{x}_{2},..., \vec{x}_{n}\right]$. For both, interpolation and differentiation in this mesh-free
 domain, it is important to find the nearest neighbors of each node $\vec{x}_{i}$. The mesh-free nodes class 
 provides this functionality by means of a kdd-tree based ordering algorithm. In our implementation
 we use the publicly available library FLANN\footnote{\href{http://www.cs.ubc.ca/research/flann/}{http://www.cs.ubc.ca/research/flann/}} \citep{muja_flann_2009}. The mesh-free nodes 
 class then provides useful features like node coordinate queries and returns node index vectors for the nearest neighbors of each node. 
 
 The second central class implements radial basis functions. The shape parameter, 
 the origin of the coordinate system and the dimensionality of the problem (1D, 2D or 3D) can bet set. 
 The user can evaluate the RBF and its derivatives at different coordinates. 
 Currently, only Gaussian RBFs are implemented but the class uses inheritance from virtual functions to generally 
 represent an RBF and allows the user to implement more general cases by providing the functional form of the RBF and its derivatives. 
 
 The combination of the mesh-free nodes and the RBF class enables the functionality, which was described in the course of this 
 section. Once the functional form of an RBF is provided, mesh-free interpolation to arbitrary evaluation points is enabled and the finite differencing weights for each evaluation point can be calculated and returned as a matrix $\mathcal{W}_{lk}$. This allows the user
 to differentiate a function $\Psi_{k} = [\Psi(\vec{x}_{1}),\Psi(\vec{x}_{2}),...,\Psi(\vec{x}_{n})]$ by 
 a simple matrix multiplication 
 \begin{equation}
 \label{equ_rbf_fd_matrix}
  D\Psi_{l} \approx \mathcal{W}_{lk}\Psi_{k}. 
 \end{equation}
Currently all differential operators $D$ up to third order are implemented in our classes.

   \section{Lensing mass reconstruction}
   \label{sec_lensing}
   We apply the RBF framework to a concrete  astrophysical application, mass reconstruction
   from gravitational lensing. 
   After a short lensing primer we show how different constraints from gravitational lensing can be combined
   in a free-form way by using a mesh-free approach.

   \subsection{Lensing primer}
   Einstein's theory of general relativity predicts the deflection
   of light rays due to gravitational potentials \citep[see e.g.][for a complete derivation]{Bartelmann2010a}. 
   By introducing the thin-lens approximation, which assumes that the distances between objects in the lensing scenario
   are much larger than the spatial extent of these objects, the lens mapping can be described by a lens equation
\begin{equation}
\label{equ_lensequation}
\vec{\beta}= \vec{\theta}-\vec{\alpha}\left(\vec{\theta}\right).
\end{equation}
This central equation describes how the 2D angular position in the source plane $\vec{\beta} = (\beta_{1},\beta_{2})$ is mapped by a deflection angle $\vec{\alpha} = (\alpha_{1},\alpha_{2})$ onto the angular coordinates $\vec{\theta} = (\theta_{1},\theta_{2})$ in the lens plane. 
The deflection angle can be related to a lensing potential
\begin{equation}
\label{equ_lensingpotential}
  \psi(\vec{\theta}):= \frac{1}{\pi}\int d^{2}\theta^{'}\frac{\Sigma(D_{\mathrm{l}}\vec{\theta})}{\Sigma_{\mathrm{cr}}}\textrm{ln}|\vec{\theta}-\vec{\theta}^{'}|,
\end{equation}
that inherits the surface-mass density of the lens $\Sigma(D_{\mathrm{d}}\vec{\theta})$.
The cosmological background model enters this equation through the critical surface mass density for lensing given by
\begin{equation}\label{critical_dens}
\Sigma_{\mathrm{cr}}=\frac{c^{2}}{4\pi G}\frac{D_{\mathsf s}}{D_{\mathrm l}D_{\mathsf{ls}}},
\end{equation}
where $c$ is the speed of light and $G$ is Newton's constant. 
The angular diameter distance between observer and lens $D_{\mathrm{l}}$, between observer and source $D_{\mathrm{s}}$, and between lens and source $D_{\mathrm{ls}}$ set the geometry of the lensing scenario. 

When introducing the edth operators \citep{Newman1962} $\partial := (\frac{\partial}{\partial\theta_{1}} +\mathrm{i}\frac{\partial}{\partial\theta_{2}})$ and $\partial^{*} := (\frac{\partial}{\partial\theta_{1}} -\mathrm{i}\frac{\partial}{\partial\theta_{2}})$, 
lensing quantities are easily  
related to the lensing potential \citep[see e.g.][and references therein]{Bartelmann2001}
\begin{align}
\label{equ_lensingquantities}
 \alpha &:= \partial\psi \qquad \qquad  &s=1 \nonumber \\
 2\gamma &:=  \partial\partial \psi  &s=2 \\
 2\kappa &:= \partial \partial^{*} \psi  &s=0 \nonumber
\end{align}
where $\alpha$ is the deflection angle, $\gamma$ is called the complex shear and the scalar quantity $\kappa$ is called convergence. The spin-parameter $s$ describes the transformation properties of each quantity under rotations of the
coordinate frame.  

The regime of weak gravitational lensing is governed by small distortions in the shape of background galaxies, which 
are observationally measured as complex ellipticities $\epsilon(\vec{\theta})$.
Once we introduce the reduced shear
\begin{equation}\label{equ_reducedshear}
  g:=\frac{\gamma}{1-\kappa},
\end{equation}
we can establish the connection between localized ellipticity averages over an ensemble of sources and the properties of the lens
\begin{equation}
 \left<\epsilon\right> = 
\left\{ \begin{aligned}
&g &\qquad &\textrm{for }|g|\leq 1 \\
&\frac{1}{g^{*}} &\qquad &\textrm{for } |g| > 1.
\end{aligned}
\right.
\end{equation}
The local averages are necessary to separate the lensing signal from random orientations due to the intrinsic ellipticity of galaxies.
For a thorough review of weak lensing and its applications we refer to \citet[][and references therein]{Bartelmann2001}
and for a discussion of systematic effects in weak lensing studies to \citet{Kitching2012}, \citet{Massey2013} and \citet{Mandelbaum2014}.

In the strong lensing regime the lens equation becomes non-linear, 
multiple images of the same source can form and shape distortions are not small any more. This leads, in some cases,
to the formation of spectacular gravitational arcs or even rings in the vicinity of a strong lens. 
The spatial extent of this regime, close to the core of the lens where densities are highest, is indicated 
by the critical curve at a given redshift. It is defined by the roots of the determinant of the lensing Jacobian
\begin{equation}
\label{equ_jacobiandeterminant}
 \det\mathcal{A} = (1-\kappa)^{2}-\gamma^{*}\gamma.
\end{equation} 

   \subsection{Combining lensing constraints in the mesh-free domain}
   \label{subsec_unstrucgrid_lensing}
   
\begin{figure}
      \includegraphics[width=.5\textwidth]{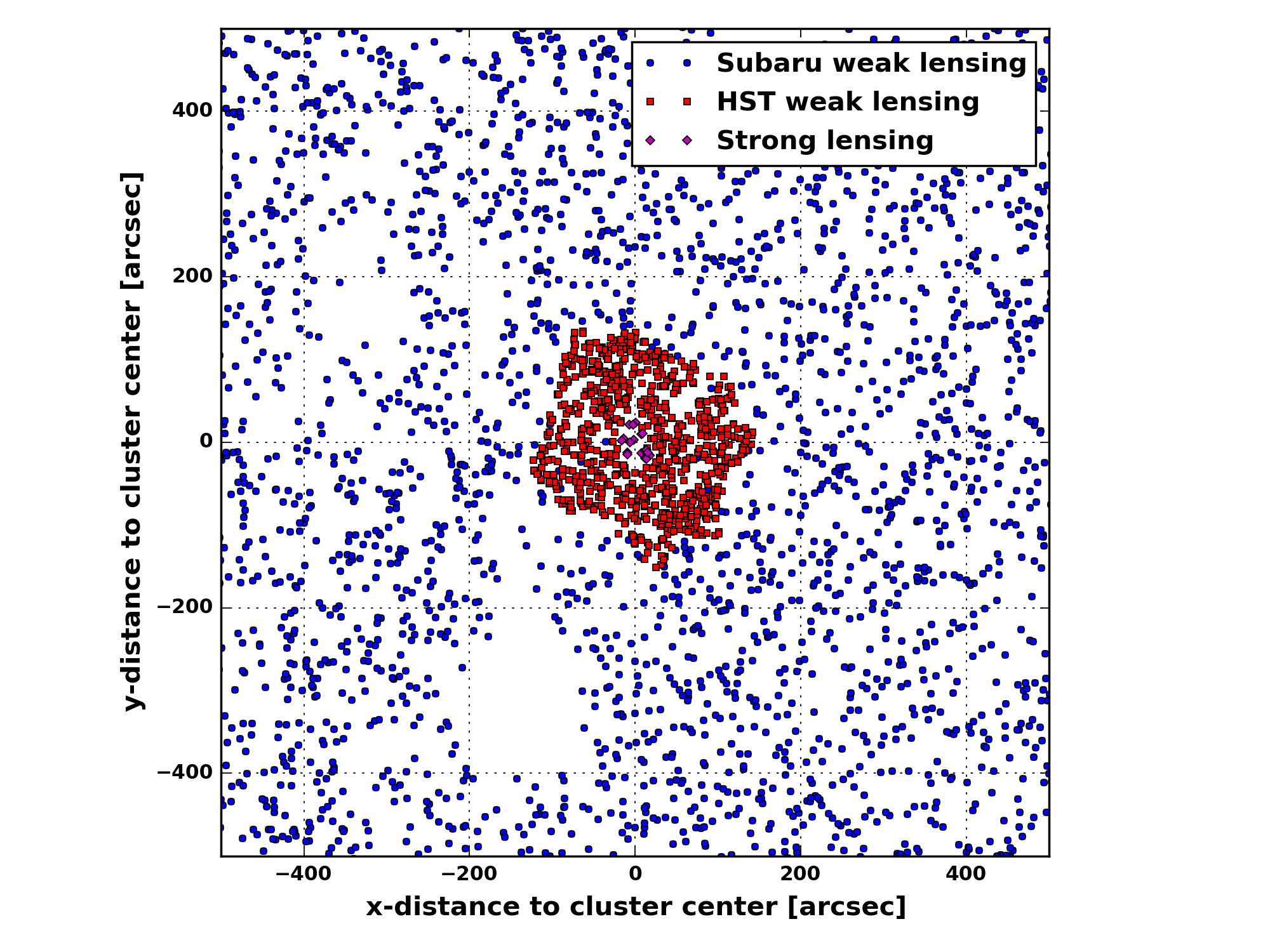}
     \caption{The distribution of lensing constraints in the galaxy cluster Abell~383 as seen by CLASH \citep{Zitrin2011e,Postman2012,Merten2015}.}
     \label{fig_lensing_observables}
    \end{figure}

   The toolkit developed in Sec.~\ref{sec_rbf} can be used to perform mesh-free lensing reconstructions. 
   That is to recover the underlying total mass distribution of a lens from weak -and strong lensing constraints.
   Eqs.~\ref{equ_lensingquantities}, \ref{equ_reducedshear} and \ref{equ_jacobiandeterminant} show that all these effects 
   can be related to spatial derivatives of the lensing potential. Since we can calculate numerical derivatives from randomly distributed nodes, 
   we construct a reconstruction method by assigning each lensing observable an evaluation point in the mesh-free domain. 
   A typical distribution of such observables is shown in Fig.~\ref{fig_lensing_observables} with the galaxy cluster Abell~383, 
   as seen by the Cluster Lensing and Supernova Survey with Hubble (CLASH) \citep{Zitrin2011e,Postman2012,Merten2015}, as an example. 
   
   Once we have chosen the number of nearest neighbors in the finite differencing (FD) stencil, we can express
   Eq.~\ref{equ_FD} as a matrix operation (Eq.~\ref{equ_rbf_fd_matrix}). In the specific case of the lensing quantities related to the 
   discretized lensing potential $\psi = \left[\psi(\vec{x}_{1}),\psi(\vec{x}_{2}),...,\psi(\vec{x}_{N})\right]$ we find
   \begin{align}
   \label{equ_pot_alpha}
    \alpha^{1,2}_{l} &= \mathcal{D}^{1,2}_{lk}\psi_{k}\\
    \label{equ_pot_gamma}
    \gamma^{1,2}_{l} &= \mathcal{G}^{1,2}_{lk}\psi_{k}\\
    \label{equ_pot_kappa}
    \kappa_{l} &= \mathcal{K}_{lk}\psi_{k}
   \end{align}
   where $\mathcal{D}$, $\mathcal{G}$ and $\mathcal{K}$ contain the FD weights for the differential operators in Eq.~\ref{equ_lensingquantities}.
   In order to recover the lensing potential from the input constraints we define a $\chi^{2}$-function, 
   which relates the lensing observations at each evaluation point
   to the lensing potential. We will define all components of this function in different lensing regimes later on.
   In order to find the lensing potential which is most likely to 
  have caused the observed lensing effects we minimize	 the $\chi^{2}$-function with respect to the potential values at each evaluation point
   \begin{equation}
   \label{equ_psichi2}
   \frac{\partial\chi^{2}(\psi_{k})}{\partial\psi_{l}}\stackrel{\text{!}}{=}0;
 \end{equation}
 where we use
 \begin{equation}
 \label{equ_pot_delta}
 \frac{\partial \mathcal{W}_{lk}\psi_{k}}{\partial\psi_{l}}=\delta_{lk}
 \end{equation}
 for any matrix representation $\mathcal{W}$ of a FD
 stencil.

   \subsubsection{Weak lensing}
   It is our goal to combine several lensing constraints into a joint reconstruction algorithm and we start
   with the weak lensing contribution. Eq.~\ref{equ_reducedshear} shows 
   that average measured ellipticities of background galaxies are directly related to the reduced shear of the lens. 
   Hence, we write the weak lensing term as
   \begin{equation}
    \label{equ_chi2WL}
    \chi^{2}_{\mathsf w}=\sum\limits_{i,j}(\left<\varepsilon\right> -g(\psi))_{i}\mathcal{C}^{-1}_{ij}(\left<\varepsilon\right> -g(\psi))_{j},
   \end{equation}
   where the indices $i,j$ run over all weak lensing evaluation points. It has been shown in 
   \citet{Bradav2005} and \citet{Merten2009} how the minimization of such a $\chi^{2}$ with respect to the discretized lensing potential
   can be written as a linear system of equations, while using Eqs.~\ref{equ_pot_gamma}, \ref{equ_pot_kappa} and Eq.~\ref{equ_pot_delta}.
   We refer the interested reader to Appendix A of \citet{Merten2009}, which shows the full derivation but we give additional information in App.~\ref{app_lse}. We have to mention that we limited ourselves to the case $|g| \leq 1$. Reliable weak lensing shape measurements
   near the very center of strong lenses are challenging and may introduce unwanted systematic effects. Henceforth, we exclude the regime where $|g| >1$ from our weak lensing analysis, but rely 
   on constraints from strong lensing in these areas. In principle, however, and given a reliable shape measurement in this regime, weak lensing reconstruction nodes with  $|g| >1$ can be included in the reconstruction scheme, 
   as e.g. shown in \citet{Bradav2005}.
   
   The covariance matrix of the weak lensing data $\mathcal{C}_{ij}$ deserves special attention. It is well-known that galaxies
   carry an intrinsic ellipticity with a standard deviation of $\sigma_{\epsilon} \sim 0.3$ \citep[e.g.][]{Chang2013}, which is not induced by lensing.
   One way of incorporating this into the $\chi^{2}$ minimization is to assume that the galaxy intrinsic ellipticities are
   uncorrelated, which results in a diagonal covariance matrix with the canonical value of $0.3^{2}$ for all its non-zero elements. 
   However, in the presence of noise, this approach leads to a very poor recovery of the lensing potential, as we will show later on.
   A different approach exploits the fact that the intrinsic ellipticity of galaxies has no preferred orientation\footnote{This statement of course ignores the effect of intrinsic alignments of galaxies in wide-field shear surveys \citep[e.g.][]{Hirata2004},
   but for our purposes of cluster lensing by individual objects, it is certainly justified.}. 
   To a given weak lensing evaluation point we do not assign the ellipticity value of a single galaxy but we perform a distance-weighted average over 
   an ensemble of nearest neighbors with respect to the point of interest. By doing so, the undirected, intrinsic ellipticity signal averages out and the 
   coherent lensing signal remains. This procedure obviously introduces correlations between the neighboring pixels which 
   were used to define the ellipticity samples. We calculate this sample covariance following Equation 15 of \citet{Merten2009} and take it into
   account by summing over the full covariance in the $\chi^{2}$-minimization. 

   \subsubsection{Strong lensing}
   One of the biggest advantages and indeed the biggest motivation 
   for a mesh-free reconstruction algorithm is the fact that the 
   distribution of evaluation points is intrinsically adaptive. This is important because different lensing constraints are 
   confined to quite different length scales, as is clearly seen in Fig.~\ref{fig_lensing_observables}. The mesh-free
   approach allows us to place evaluation points where data is available. In the case of weak lensing,
   this spans the entire field of the lens, with usually increased resolution towards the center when high-quality 
   data from e.g. the Hubble Space Telescope (HST) is available. Strong lensing is confined to the very core 
   of the lens and allows for a very finely grained recovery of the lensing potential if many strong lensing features are observable.

One constraint related to strong lensing are multiple-image systems. Our $\chi^{2}$-minimization 
term is similar to \citet{Bradav2005} but differs in some details. The general idea is based on the fact that different
images $i$ of the same multiple-image system should be mapped back to the same position in the source plane. Therefore, we write
a $\chi^{2}$-term
 \begin{equation}
  \chi^{2}_{\mathsf m} = \sum\limits_{i = 1}^{N_{s}} \left(\frac{\beta_{i}(\psi)-\left<\beta\right>(\psi)}{\sigma_{i,\mathsf{m}}}\right)^{2} 
  \label{msystemchi2}
\end{equation}
where $\beta_{i}$ is the source-plane position of each image of the system and
 \begin{equation}
 \left<\beta\right> = \frac{1}{N_{s}}\sum\limits_{i = 1}^{N_{s}}\beta_{i}
\label{betadef}
\end{equation}
is the average source-plane position of all images in the system. The total number of images in a single multiple-image system is $N_{s}$.
In the last two equations we can use Eq.~\ref{equ_lensequation} to replace
the source-plane position $\vec{\beta}$ with the observed lens-plane position $\vec{\theta}$ and the deflection angle $\vec{\alpha}$,
which carries the wanted dependence on the lensing potential
\begin{equation}
\label{equ_chi2MSYS}
 \chi^{2}_{\mathsf m} = \sum\limits_{i = 1}^{N_{s}}\frac{1}{\sigma_{s}^{2}} \left(\vec{\theta}_{i}-\vec{\alpha}_{i}(\psi)-\frac{1}{N_{s}}\sum\limits_{j=1}^{N_{s}}\left(\vec{\theta}_{j}-\vec{\alpha}_{j}(\psi)\right)\right)^{2}.
\end{equation}
Here, $\sigma_{s}$ is the tolerated positional error on each image position in the source plane. 
The result of minimizing this $\chi^{2}$-function with respect to the lensing potential is shown in App.~\ref{app_lse}.
In the presence of more than one multiple-image system, one adds a $\chi^{2}$-term for each system, respectively.

      The second strong lensing constraint are estimates on the position of the
   critical curve of the lens. This  has been discussed in
   detail in e.g. \citep{Jullo2007,Merten2009,Merten2011,Merten2015}. The corresponding $\chi^{2}$-term enforces the lensing Jacobian to vanish for pixels which are assigned to be part of the critical curve and which are indicated 
   in the following by a pixel index $c$
   \begin{equation}
    \chi^{2}_{\mathrm s}(\psi)=\sum\limits_{c=1}^{N_{c}}\frac{|\det\mathcal{A}(\psi)|^{2}_{c}}{\sigma^{2}_{c,\mathrm{s}}}.
   \label{equ_chi2SL}
   \end{equation}
   The total number of these estimators is $N_{c}$ and the error $\sigma^{2}_{\mathrm{s}}$ derives from a positional error that is assigned to the critical curve estimator.
   We approximate it via
   \begin{equation}
   \sigma_{\mathsf s}\approx \left.\frac{\partial\det\mathcal{A}}{\partial \theta}\right|_{\theta_{\mathsf c}}\delta\theta\approx \frac{\delta\theta}{\theta_{\mathsf E}}, 
   \label{equ_chi2strong_sigma}
   \end{equation}
where $\theta_{\mathsf E}$ is an estimate for the Einstein radius of the lens. The minimization of the
$\chi^{2}$-function related to this constraint can also be related to a linear system of equations using Eqs.~\ref{equ_pot_gamma},\ref{equ_pot_kappa} and Eq.~\ref{equ_pot_delta}
and we again leave the actual calculation to Appendix A of \citet{Merten2009}.
When using critical line estimators as a constraint, one has to keep in mind that they are not a direct observable in a lensing scenario. The advantages and caveats when using 
these constraints, together with the accuracy of lensing reconstructions has been discussed earlier \citep[e.g.][]{Cacciato2006,Merten2009,Meneghetti2010a, Merten2015}. We still include this 
constraint in this paper since it is a useful feature in a lensing reconstruction method. One can for example incorporate constraints from another pure strong lensing method into the current
reconstruction via these estimators.

\subsubsection{Implementation}
In order to find the lensing potential which causes the joint observations of all weak -and strong lensing terms we sum
over the independent contributions of the $\chi^{2}$ function and find a single linear system of equations. The solution
to this linear system is the mesh-free representation of the reconstructed lensing potential, from which all other quantities
of interest can be derived using Eqs.~\ref{equ_pot_alpha}, \ref{equ_pot_gamma} and \ref{equ_pot_kappa}. 

In order to guarantee a smooth reconstruction in the presence of noisy weak lensing data we introduce an outer-level iteration,
following the scheme of \citet{Bradav2005}, which is also used in \citet{Merten2009}. We define a regularization term in the
$\chi^{2}$-function which controls the reconstruction in such a way that the result will not diverge strongly from a well-defined
regularization condition. In our case, this condition is set by pre-defined
convergence $\kappa_{\mathrm{reg}}$ and shear $\gamma_{\mathrm{reg}}$ solutions
\begin{align}
 \label{equ_convregularisation}
 \chi^{2}_{\mathrm{c\_reg}}&=\sum\limits_{i=1}^{N}\eta^{c}_{i}\left(\kappa_{i}(\psi)-\kappa_{\mathrm{reg}}\right)^{2}\\
 \label{equ_shearregularisation}
 \chi^{2}_{\mathrm{s\_reg}}&=\sum\limits_{i=1}^{N}\eta^{s}_{i}\left(\gamma_{i}(\psi)-\gamma_{\mathrm{reg}}\right)^{2}.	
\end{align}

In the summation above, $i$ runs over all evaluation points and it should be noted that in this
implementation the strength of the regularization $\eta$ can be set for each evaluation point individually and can be set to different values for convergence and shear regularization, respectively. 
The idea of the outer-level iteration is then to start with only
few weak lensing evaluation points and to average ellipticities of a large sample of weak lensing sources for each of these nodes. 
This results in a coarse but almost shape-noise free reconstruction. In the following steps, the number of nodes is continuously
increased, resulting in smaller ellipticity samples but relying on a reconstruction that is convergence- and shear-regularized on the result of the former 
reconstruction step. This outer-level iteration effectively reduces the noise-level in the reconstruction, as shown in \citet{Bradav2005} and
\citet{Merten2009}, and as we will explore in our accuracy tests later on. 

Ultimately, we are solving the linear system of equations which is calculated from the minimization of the $\chi^{2}$-function
\begin{equation}
 \chi^{2}(\psi) = \chi^{2}_{w}(\psi)+\chi^{2}_{s}(\psi)+\chi^{2}_{m}(\psi)+\chi^{2}_{\mathrm{s\_reg}}(\psi)+\chi^{2}_{\mathrm{c\_reg}}.
\end{equation}
One notes that the weak lensing and the critical-line
estimator term contain non-linear contributions. We account for this by introducing an inner-level iteration, following
the scheme of \citet{Schneider1995}. During each inner-level reconstruction iteration, non-linear terms in the $\chi^{2}$-function are isolated and held constant
in order to solve the linear system of equations. New estimates for convergence and shear are calculated from this solution and new approximations for the non-linear terms  
are inserted as constants into the linear system of equations. This iteration converges after 2-5 reconstruction steps. 
 
\begin{figure}
      \includegraphics[width=.45\textwidth]{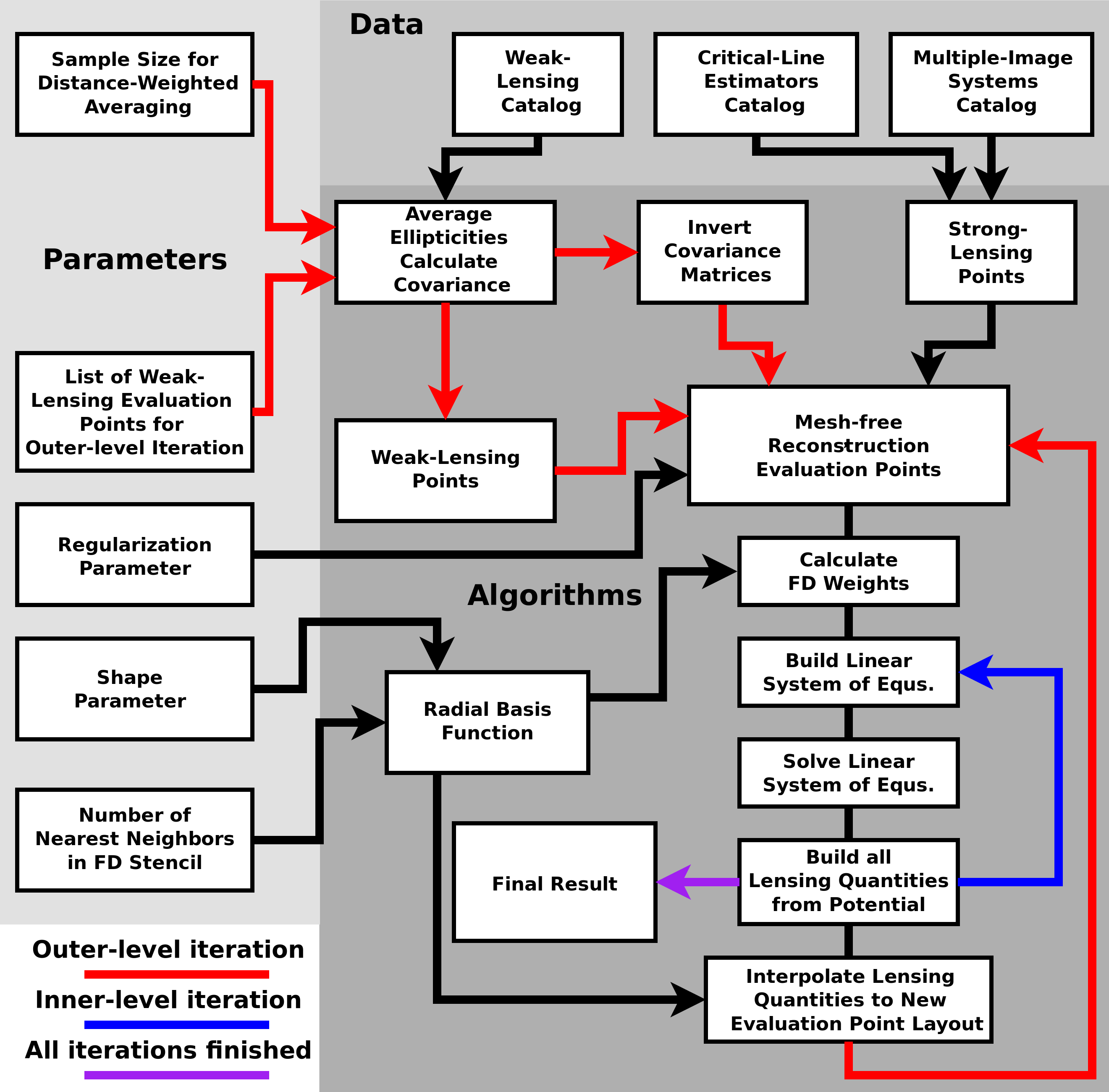}
     \caption{A flowchart  of the reconstruction process using weak and strong lensing constraints.}
     \label{fig_flowchart}
    \end{figure}

A complete flowchart of the reconstruction algorithm is shown in Fig.~\ref{fig_flowchart}. Initially, the weak lensing 
catalog is read and depending on the stage of the outer-level iteration, ellipticity values are averaged to define the weak lensing evaluation points. 
These are then combined with the strong lensing evaluation points, which directly derive from the critical-line estimator and multiple-image
system catalogs. For all these nodes, the finite-differencing stencils are calculated using RBFs
and the reconstruction is performed, including the inner-level iteration. This procedure is repeated with increasing resolution
until all outer-level iteration steps are completed. Each reconstructed result is interpolated
to the next larger mesh-free outer-level configuration using RBFs. The convergence and shear maps derived from this interpolation serve as the regularization template
for the next step. The very first regularization template depends on the reconstruction and the field of view of the data but, in
most cases, a flat and zero convergence and shear template suffices. However, a more sophisticated choice for the initial prior is also 
possible resulting in more complicated initial convergence and shear regularization templates. 

   \section{Accuracy tests with mock lenses}
   \label{sec_testing}
   We use two different test scenarios to check the implementation and performance of our reconstruction method. A simple toy model lens provides the framework to check the basic implementation in various stages. 
   A simple toy model based test is followed by a much more realistic lensing scenario, which is based on a full ray-tracing simulation of a simulated cluster-sized halo. There we also mimic several observational effects and sources of noise.
   
   \subsection{Toy model lens}
   \label{sec_testing_toy_model}
\begin{figure}
      \includegraphics[width=.5\textwidth]{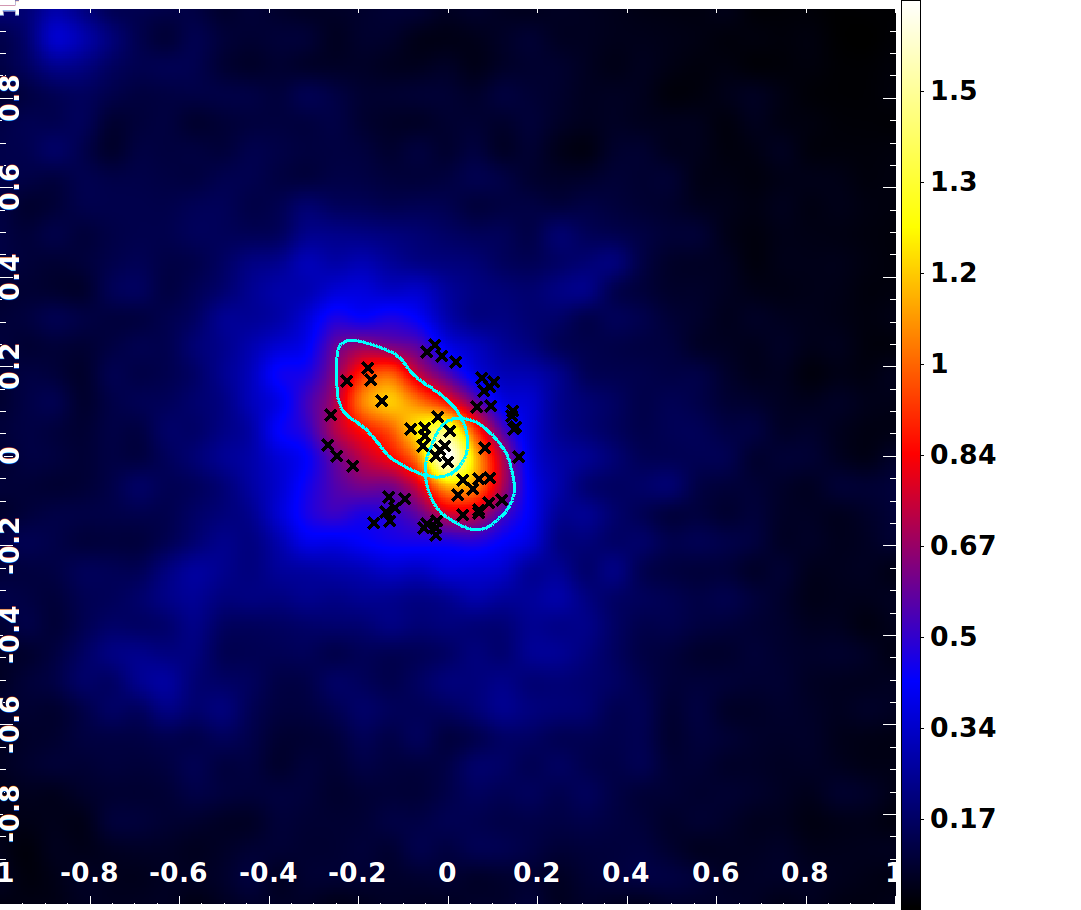}
     \caption{The convergence map of our mock lens on the unit square. The cyan line in the center of the image is the 
     critical line of the lens for a fiducial redshift of $z_{\mathrm{s}}=1.0$. The black crosses indicate the positions of multiple images
     that we use to reconstruct the lens.}
     \label{fig_mock_lens}
    \end{figure}

   We use a numerically simulated, cluster-sized lens to to provide a simple proof of concept for the applicability of our theoretical concept.
   This mock lens is described in more detail in \citet{Bartelmann1998} and was already used for accuracy tests in
   \citet{Cacciato2006} and \citet{Merten2009}. The surface-mass
   density map of this lens is shown in Fig.~\ref{fig_mock_lens}. The side length of the field of view is 
   5 Mpc/h or $18\arcmin$ at the lens' redshift of $z=0.35$. The Einstein radius of the lens is $\theta_{\mathrm{E}}\sim30\arcsec$ for
   a source redshift of $z_{\mathrm{s}}=1.0$. In the following, especially in the figures of this section, 
   we will scale these distances to dimensionless coordinates
   by mapping them into the unit-square with side length 2.
   From the known deflection angle fields, we sample
   the following lensing catalogs to serve as input for our test reconstructions:
   \begin{itemize}
    \item 9000 complex shear values at random positions in the field. This refers to a background-galaxy density of $\sim 25~\mathrm{arcmin}^{-1}$.
    \item The same catalog of 9000 weak lensing shear values but with an added shape-noise component. 
    This noise is sampled randomly in each shear component from a Gaussian distribution with a standard deviation of 0.2 (compare Sec. ~\ref{subsec_unstrucgrid_lensing}).     
    \item 55 multiple images, by randomly placing 5 point sources inside the inner caustic and 10 point sources in between the inner and the outer caustic
    of the lens. 
    \item 20 critical line estimators by randomly sampling points in the field for which the determinant of the lensing Jacobian (Eq.~\ref{equ_jacobiandeterminant})
    vanishes within the limits of the numerical precision. 
   \end{itemize}
   All these lensing constraints are placed at a fiducial redshift of $z_{\mathrm{s}}=1.0$.

   In the following we will investigate how well we can reconstruct the underlying mass distribution from weak lensing alone, where
    we investigate the influence of important complications to such analyses.
   
   As a word of caution we want to stress that these tests can only be seen as a proof of concept. More thorough and realistic tests with fully realistic lensing scenarios \citep[e.g][]{Meneghetti2010a} 
   have to follow, together with a full comparison to other mass reconstruction methods. Furthermore, the readiness of the method to be applied to real astronomical data has to be shown in such a follow-up analysis.

  \subsubsection{Weak lensing tests}
\begin{figure}
      \includegraphics[width=.5\textwidth]{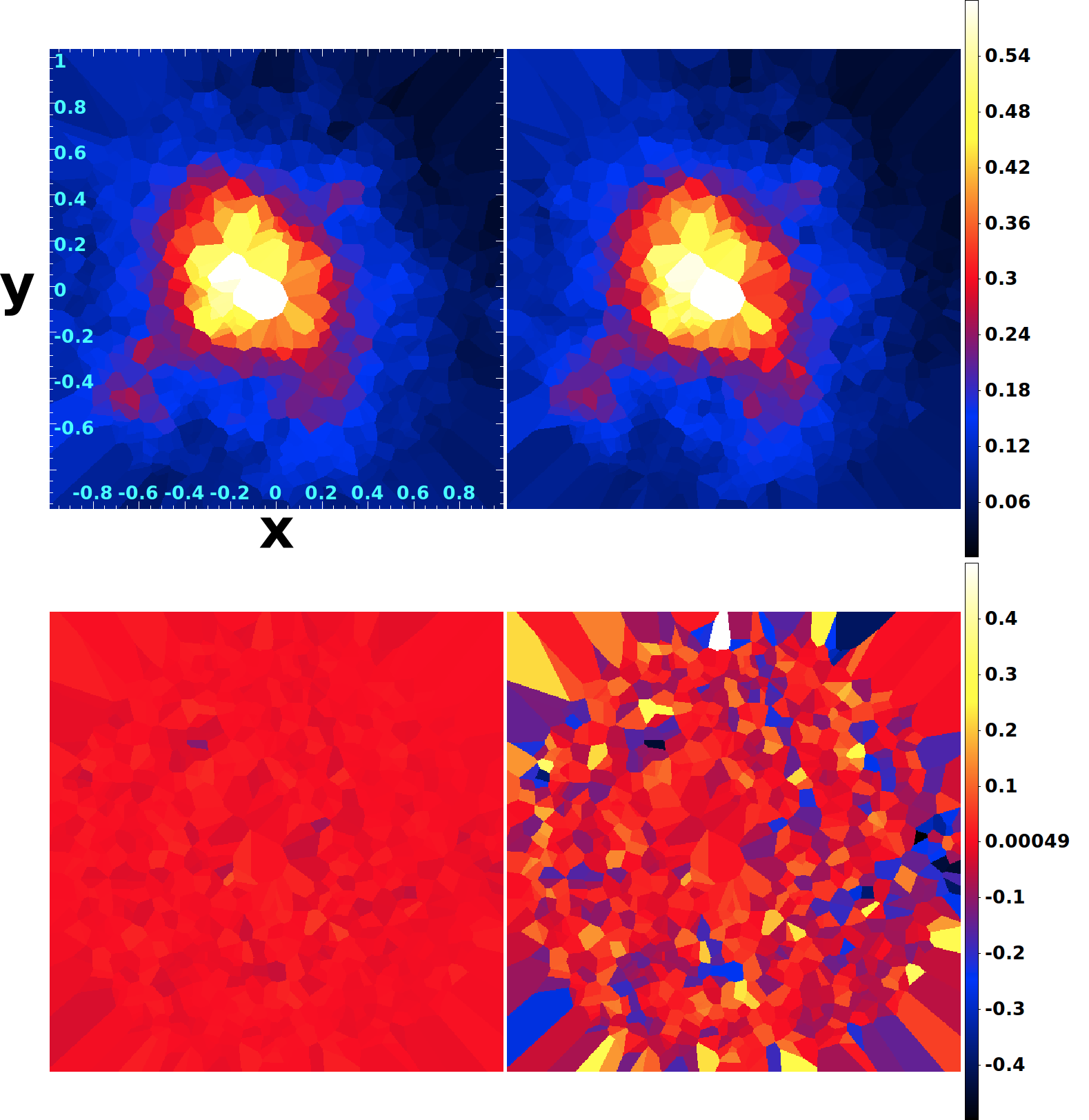}
     \caption{The result of the weak lensing only reconstruction of the mock lens as a Voronoi representation.
     This reconstruction is performed without any noise contribution in the weak lensing input data.
     The left panel shows the convergence
     map of the reconstruction. The right panel shows the relative error in the reconstructed convergence when compared to 
     the true input at the evaluation points of the reconstruction.}
     \label{fig_WL_rec_noNoise}
    \end{figure}

  We perform a pure weak lensing reconstruction of the mock lens using the shear catalog without any shape noise contribution first. From the 9000 ideal shear values we randomly
  pick 900 to serve as evaluation points of the mesh-free reconstruction. Since the data contains no noise component, no outer-level iteration is needed,
  but we still assign the fiducial value $\sigma_{\epsilon}=0.3$ to each weak lensing constraint.
  However, we use the reduced shear as weak lensing input which demands the inner-level iteration to compensate
  for the non-linear contributions to the $\chi^{2}$-minimization. In order to correct for the mass-sheet degeneracy we force the very 
  upper-right corner of the reconstructed region to approach a convergence value of zero. We present the result in Fig.~\ref{fig_WL_rec_noNoise} where 
  the top panel shows the real convergence map of the mock lens on the left and the reconstructed convergence map on the right. 
  Both maps follow the same resolution based on the 900 weak lensing evaluation points. 
  The bottom panels show the absolute difference between the reconstructed
  convergence map and the real one on the left and the relative difference to the real map on the right. The general agreement
  is quite striking, as it should be under these perfect test conditions. The average absolute difference
  between the maps is -0.002 and the average relative difference is -0.02. We performed this over-simplified reconstruction to serve as a benchmark for our additional 
  reconstructions which include typical complications for weak lensing analyses.

\begin{figure}
      \includegraphics[width=.5\textwidth]{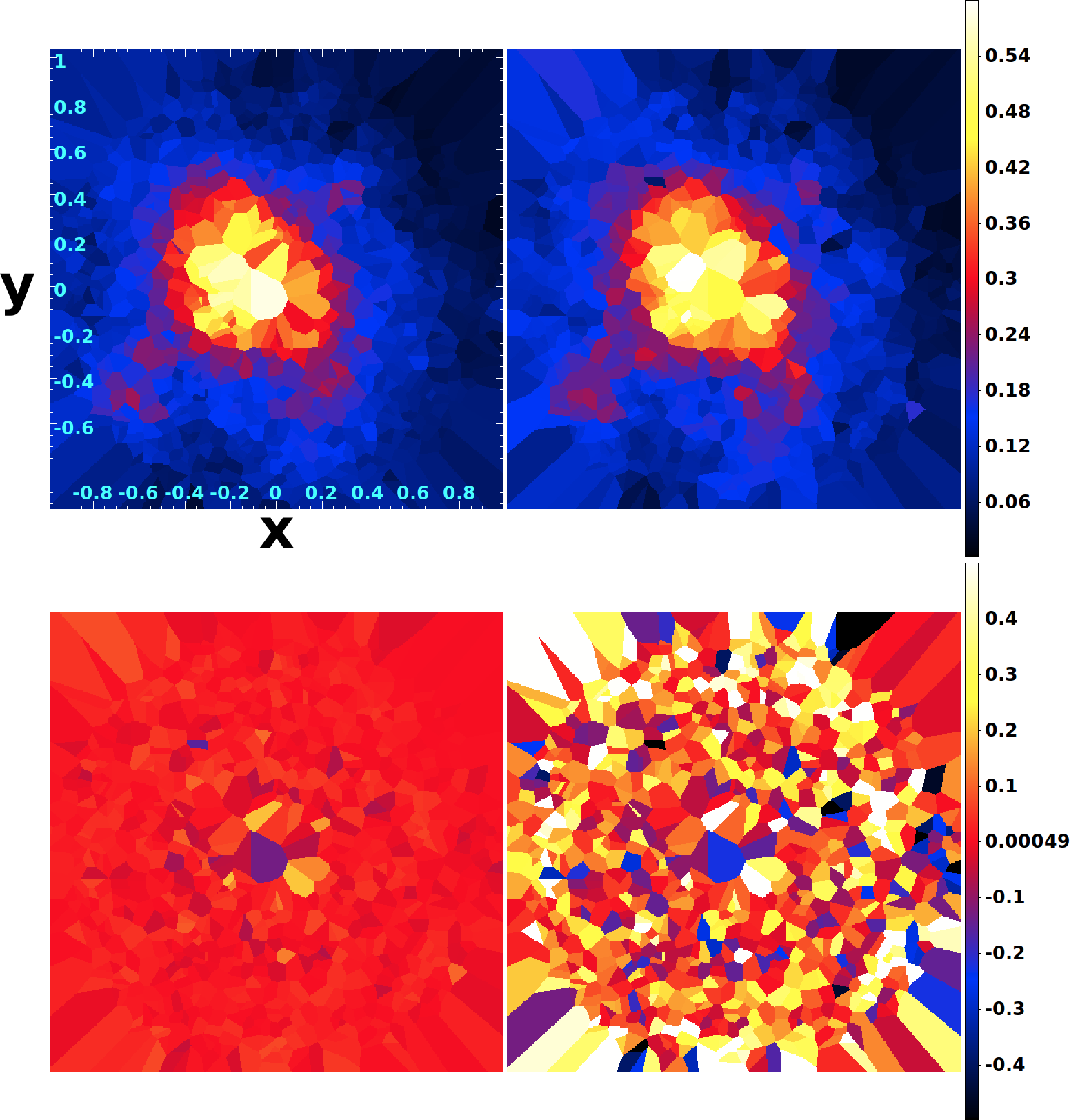}
     \caption{The result of the weak lensing only reconstruction of the mock lens as a Voronoi representation. 
     The difference to Fig.~\ref{fig_WL_rec_noNoise} is the realistic redshift distribution of the sources in the weak lensing input catalog. The reconstruction assumes the sources to sit at a single redshift, which introduces a significant bias.}
     \label{fig_WL_rec_z}
    \end{figure}

\begin{figure}
      \includegraphics[width=.5\textwidth]{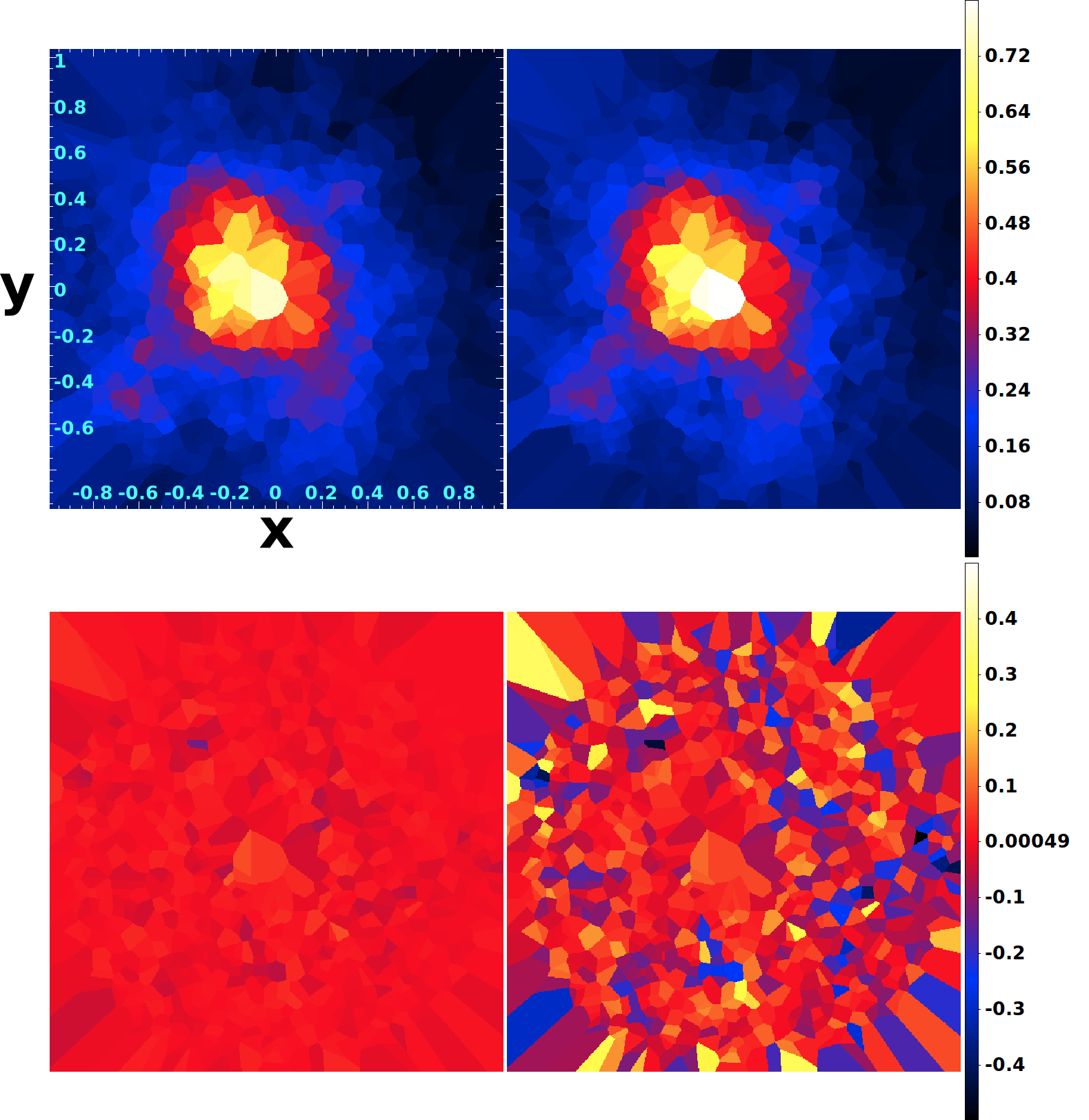}
     \caption{The result of the weak lensing only reconstruction of the mock lens as a Voronoi representation.
     This reconstruction uses the same input catalog as Fig.~\ref{fig_WL_rec_z} but accounts for the redshift distribution in the reconstruction. The convergence map are now scaled to a fiducial redshift of $\infty$ and 
     the initial accuracy level seen in Fig.~\ref{fig_WL_rec_noNoise} is now restored.}
     \label{fig_WL_rec_z_corr}
    \end{figure}

    To make the scenario more realistic we now distribute the weak lensing sources according to the photometric redshift distribution of background sources in a 
    a real galaxy cluster\footnote{Abell 209 from the CLASH survey, compare \citet{Merten2015}.}. Once we reconstruct this input catalog while wrongly assuming that all 
    sources are still placed at a single redshift we find the result presented in Fig.~\ref{fig_WL_rec_z}. The wrong assumption introduces a significant source of error, 
    pushing the average absolute difference in the maps to 0.01 and the average relative difference to 0.12. However, our method is able to deal with different redshifts for 
    each reconstruction node by scaling each node to a common fiducial redshift (see e.g. Eqs.~8,~12 in \citet{Merten2009}). Once we incorporate the knowledge about the photometric
    redshifts we find the result shown in Fig.~\ref{fig_WL_rec_z_corr}, which reduces the average, absolute difference in the maps back to -0.002 and the 
    average, relative error to -0.019. Please note that both, the real and the reconstructed convergence map are now scaled to a fiducial redshift of $\infty$.
    
\begin{figure}
      \includegraphics[width=.5\textwidth]{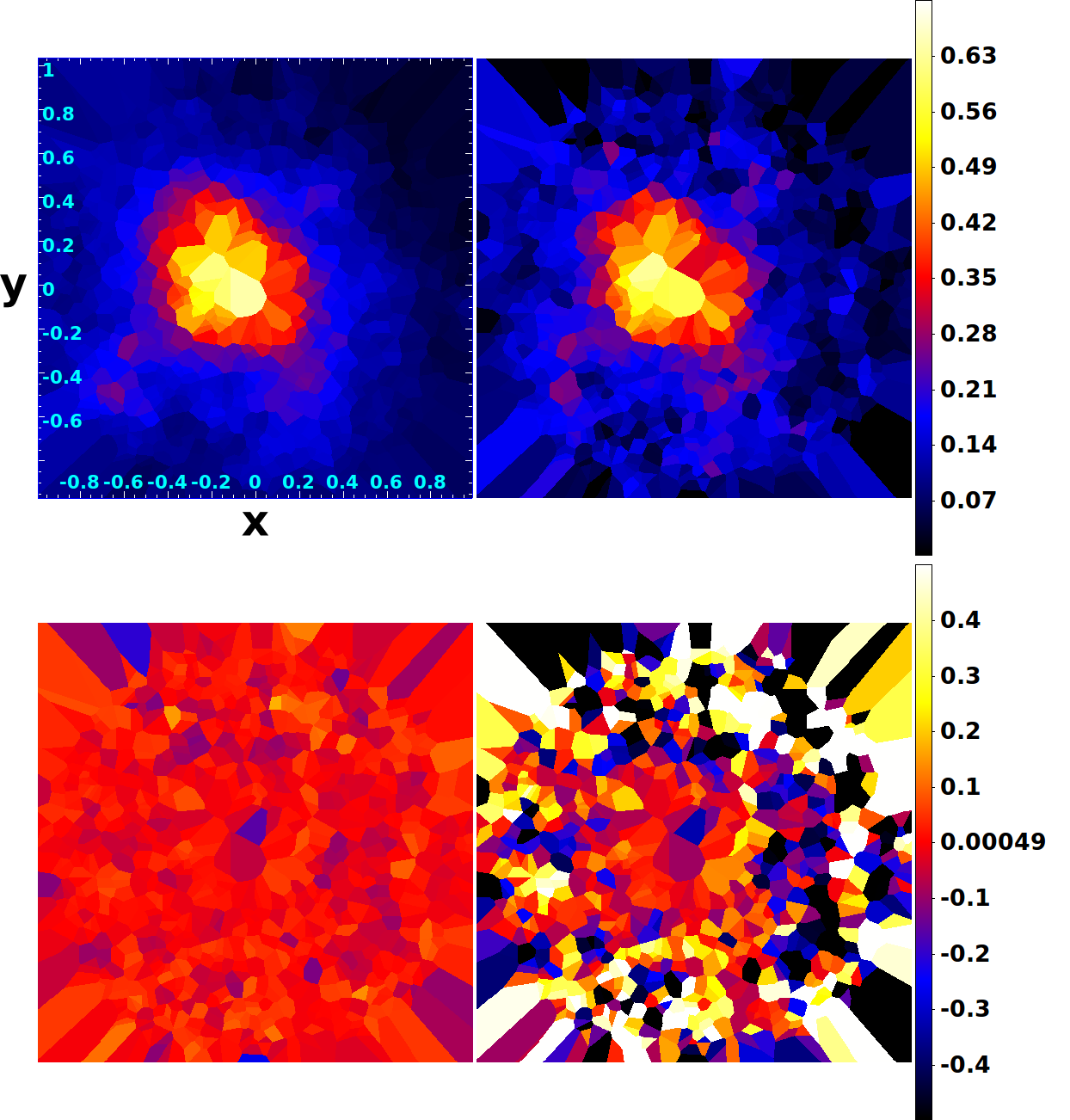}
     \caption{The result of the weak lensing only reconstruction of the mock lens as a Voronoi representation.
     This reconstruction is performed with a realistic noise contribution in the weak lensing input data.
     The left panel shows the convergence
     map of the reconstruction. The right panel shows the relative error in the reconstructed convergence when compared to 
     the true input at the evaluation points of the reconstruction.}
     \label{fig_WL_rec_Noise_iter}
    \end{figure}

    We test for another important complication by using a reduced shear catalog which contains shape noise. The result
    is shown in Fig.~\ref{fig_WL_rec_Noise_iter}. The outer-level iteration scheme, which is described
    in Sec.~\ref{subsec_num_implementation}, is implemented to minimize the effect of shape noise on the reconstruction. We define six different refinement levels by starting
    with 150 weak lensing evaluation points and gradually add 150 more points until we reach the target resolution
    of 900 nodes. The weak lensing catalog for this reconstruction is the full ellipticity sample, containing 
    9000 measurements and shape noise. For each of these six outer-level iterations 
    we perform a distant-weighted average over 80, 50, 30, 22, 18 and 15 weak lensing shear values, respectively, in order to extract the lensing
    signal from the noisy data. For the first step with 150 reconstruction cells we regularize on a flat and zero convergence and
    shear field as initial step. Later on we regularize on the interpolated results of the former, more smooth reconstruction step
    to avoid overfitting. For each outer-level resolution we perform three inner-level iterations, which is enough for the reconstruction to converge. 
    The average absolute error in the convergence is $-0.005$ and the relative 
    error drops to a value of $-0.04$. The maximum absolute and relative errors are $-0.27$ and $2.44$, respectively, which 
    are found in the outskirts of the field where the lensing signal is weakest and the shear values are largely dominated by shape noise.

   \subsubsection{Weak and strong lensing tests}
\begin{figure}
      \includegraphics[width=.5\textwidth]{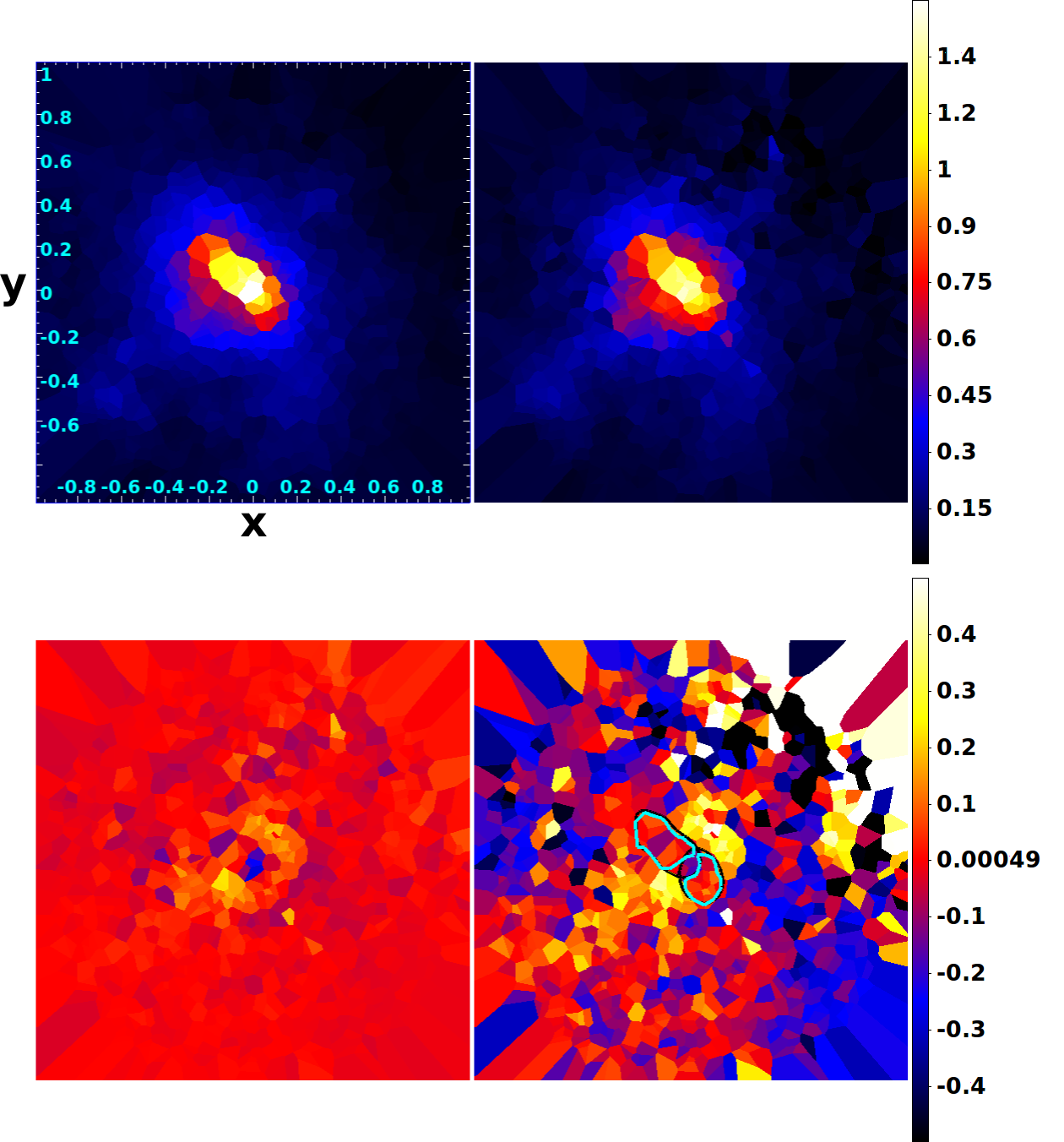}
     \caption{The Voronoi representation of the joint weak -and strong lensing reconstruction using 900 reduced shear values and
     55 multiple-image system. The top panels show the convergence of the real lens on the left and the reconstructed convergence
     map on the right. Both maps follow a resolution which is defined by the 955 input constraints. Shown in the bottom panels
     are differences between the real and reconstructed maps in terms of an absolute error on the left and a relative error with the
     real map as a reference on the right. Also shown in the bottom right panel is the reconstructed critical line in cyan and the 
     real critical line for a source redshift of 1.0 in black. The black line can barely be seen since it is well overlaid by the
     reconstructed line.}
     \label{fig_SL_WL_MS_rec_noNoise}
    \end{figure}

\begin{figure}
      \includegraphics[width=.5\textwidth]{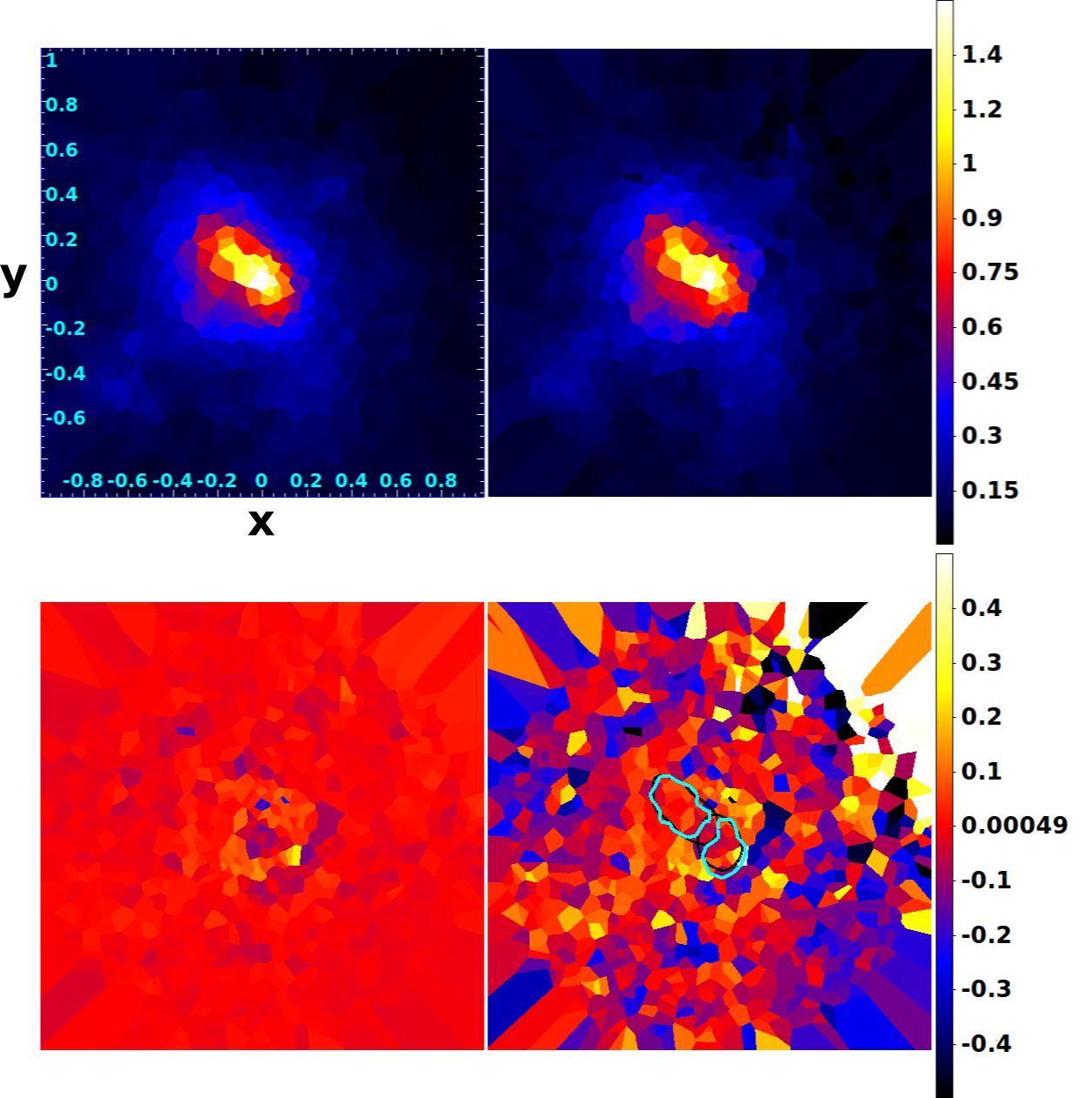}
     \caption{This figure is identical to Fig.~\ref{fig_SL_WL_MS_rec_noNoise} but uses in the underlying reconstruction additional
     20 constraints on the position of the critical line of the cluster.}
     \label{fig_SL_WL_MS_CC_rec_noNoise}
    \end{figure}

   This last effort of the testing program brings the pieces together and performs combined weak and strong lensing 
   reconstructions.
   We again draw 900 ellipticity measurements from the shape-noise free weak lensing catalog and also use the 55 multiple-image systems
   in a first joint reconstruction. The result of this reconstruction is shown in Fig.~\ref{fig_SL_WL_MS_rec_noNoise} and it is immediately
   obvious that, while the field size is identical to the weak lensing only reconstruction, the resolution in the central area
   of the lens is increased drastically due to the additional evaluation points defined by the positions of the multiple images. 
   The reconstruction yields an average absolute convergence error of $-0.007$ and the average of the relative error
   is $0.004$. The maximum absolute and relative convergence errors are $-0.29$ and $-2.4$, respectively. In the bottom
   right panel of Fig.~\ref{fig_SL_WL_MS_rec_noNoise} we also show the reconstructed critical curve of the lens and compare it with
   the real critical curve for our fiducial redshift of $z_{\mathrm{s}}=1.0$. Only small deviations between the two curves are present and the fact that the 
   reconstructed critical curve gets split into two parts is only due to the limited number of evaluation points. In general, the accuracy
   of the reconstruction is at the 5--10\% level, with the clear exception of areas just outside the critical curve where the reconstruction
   overestimates the convergence by 15--20\% especially in the areas around the $\left(x,y\right)=\left(0.0,-0.2\right)$ and the $\left(0.2,0.2\right)$
   coordinate. We overcome this shortcoming by making use of the one feature in the reconstruction that we have not used, yet.
   We add the additional 20 sample points of the critical curve of the cluster. We show the reconstruction that adds these constraints
   in Fig.~\ref{fig_SL_WL_MS_CC_rec_noNoise} and again find an excellent reconstruction but with much reduced 
   inaccuracies around the critical curve of the cluster. The average absolute and relative error is now 
   $-0.005$ and $-0.009$, respectively. The maximum absolute error is $-0.25$ and the maximum relative error is $1.75$. The recovery
   of the critical curve is still excellent as we also show in the bottom right panel of Fig.~\ref{fig_SL_WL_MS_CC_rec_noNoise}.
   
    \subsection{Realistic ray-tracing scenario}
    \label{sec_ray}
   We now move to a much more realistic lensing scenario, created with the \texttt{SkyLens} pipeline \citep{Meneghetti2008}. The lens in this ray-tracing approach is provided 
    by one 2D deflection angle map of the cluster \textit{g5699754\_G\_79235}, which is part of the simulated cluster suite described in \citet{Bonafede2011} and \citet{Fabjan2011}.
    The simulated field-of-view is 0.5 deg on a side and the corresponding surface-mass density map along our chosen line-of-sight can be seen in Fig.~\ref{fig_RAY_MAPS}.
    Following the example of \citet{Meneghetti2010a} we create the following lensing constraints in the field:
    
    \begin{itemize}
     \item 22752 galaxies with weak lensing shape measurements, which corresponds to a galaxy density of $\sim$ 25 $\textrm{arcmin}^{-1}$. In contrast to the toy model case of Sec.~\ref{sec_testing_toy_model}, these galaxies are not randomly positioned any more, but
      their spatial distribution follows a realistic, cosmological clustering scheme and is affected by shifts due to lensing.  
      More importantly, their intrinsic shape and redshift distribution follows real observations and their shape measurement is affected by real systematic effects \citep[see][for details about the ray-tracing method]{Meneghetti2008,Meneghetti2010a}.
     \item Ten multiple image systems, three of which produce five images, while the rest produces three. The redshifts of these systems cover the range from 0.970 to 3.636 in roughly equal steps.
    \end{itemize}
    
   \subsubsection{Reconstruction and error estimation}
\begin{figure}
      \includegraphics[width=.5\textwidth]{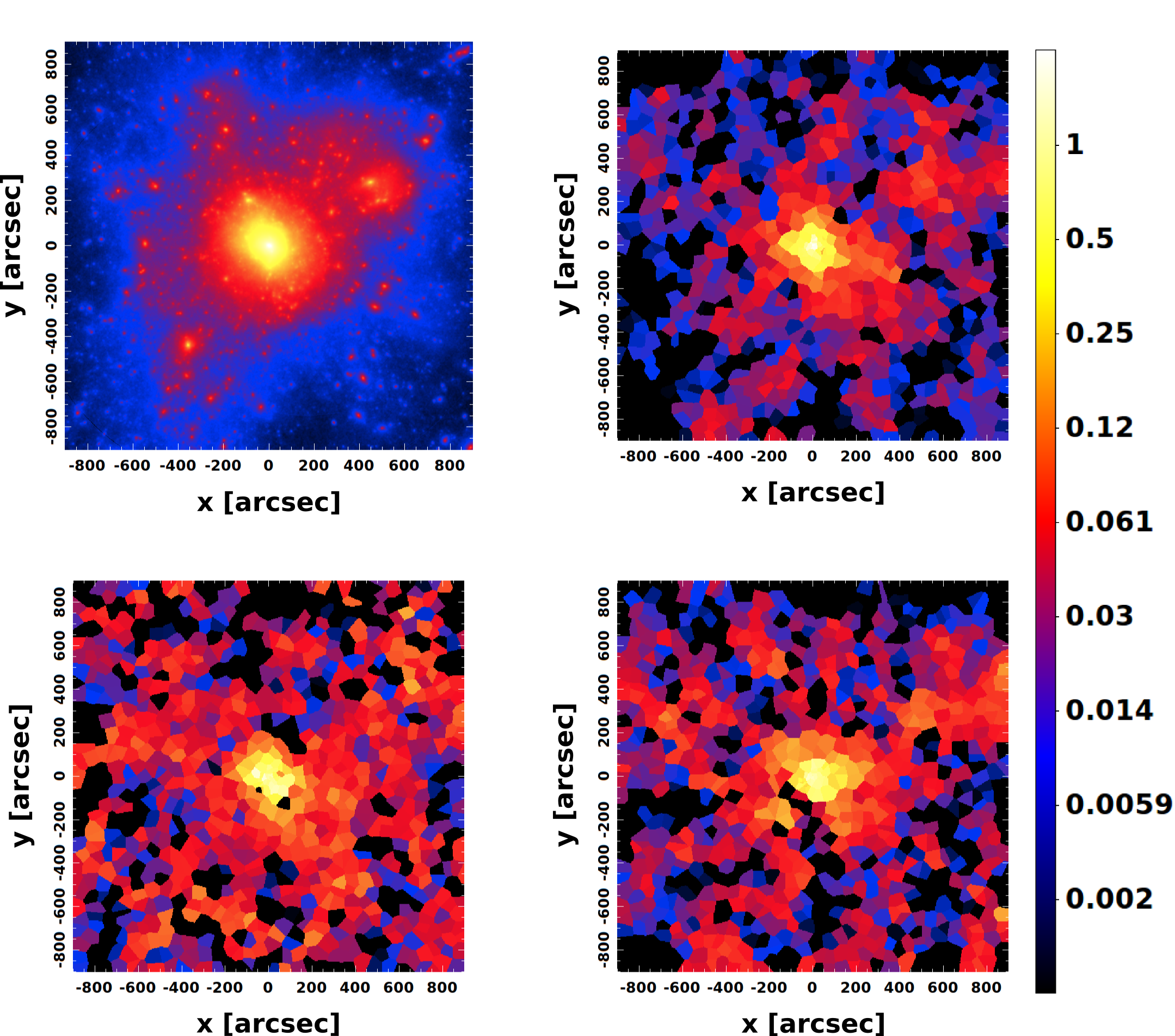}
     \caption{The real and reconstructed realistic lensing scenario. The \textit{top left panel} shows the real convergence map of the simulated lens for a source redshift of $z_{s}=2.0$. The total field size is 0.5~deg and corresponds to
      a physical scale of 7.04 Mpc. The \textit{top right panel} shows he convergence map of our fiducial reconstruction scaled to the same source redshift. The two \textit{bottom panels} show two randomly chosen reconstruction
      realizations based on input catalog resampling.}
     \label{fig_RAY_MAPS}
    \end{figure}

    We apply the methodology outlined in Sec.~\ref{sec_testing_toy_model} to the catalogs produced by the realistic lensing scenario. In order to deal with the effects of shape-noise in the weak lensing catalog and to avoid noise overfitting 
    we start the reconstruction with 200 weak lensing nodes, homogeneously covering the full field. The center of each node is chosen to be the shape measurement in the vicinity with the largest shear inverse variance.  
    Around all centers we perform a weighted average of the 100 nearest neighbors in the shear catalog. In the strong lensing regime we assign a reconstruction node to each of the 36 multiple images. We iteratively 
     increase the resolution from 236 to 1036 reconstruction nodes in steps of 200, while in each step we decrease the number of averaged shears by a factor of 0.7. Each iteration keeps memory of the earlier step through regularization with a strength which is chosen to match the weak lensing weight set by 
    covariance matrix in Eq.~\ref{equ_chi2WL}. In each of these outer level iterations we perform three inner level iterations. 
    In the last outer level iteration step, our unstructured distribution of reconstruction nodes results in a minimum distance between neighboring nodes of 1.1 arcsec and a maximum distance of 78.7 arcsec. 
    The result of this reconstruction can be seen in Fig.~\ref{fig_RAY_MAPS}, where we show the real convergence distribution of the simulated cluster next to our reconstruction. 
    One can clearly see that the reconstruction suffers from the effect of shape noise but shows a qualitatively good agreement with the real matter distribution.
    
    In order to assign realistic error bars to our reconstruction we resample our input catalogs \citep[e.g.][]{2000,Bradavc2005,Meneghetti2010a,Merten2015}. We create 250 reconstruction realizations by bootstrapping the weak lensing input catalog and by uniformly sampling a redshift
    range for each weak lensing and strong lensing constraint. We define the allowed range within 10\% below and above a constraint's real redshift. We show two examples for the resampled realizations 
    in Fig.~\ref{fig_RAY_MAPS}. The total sample of 250 realizations builds the base to determine the errors in the following analyses. In the following we refer to the original, not resampled, reconstruction as the fiducial model. The total number of 
    resamplings is mostly constrained by runtime considerations. However, we did confirm that the size of the relevant error bars is stable, when compared to smaller (e.g. 200 resamplings) sets of realizations.
    
    \subsubsection{Mass profile and NFW parameter recovery}
\begin{figure}
      \includegraphics[width=.5\textwidth]{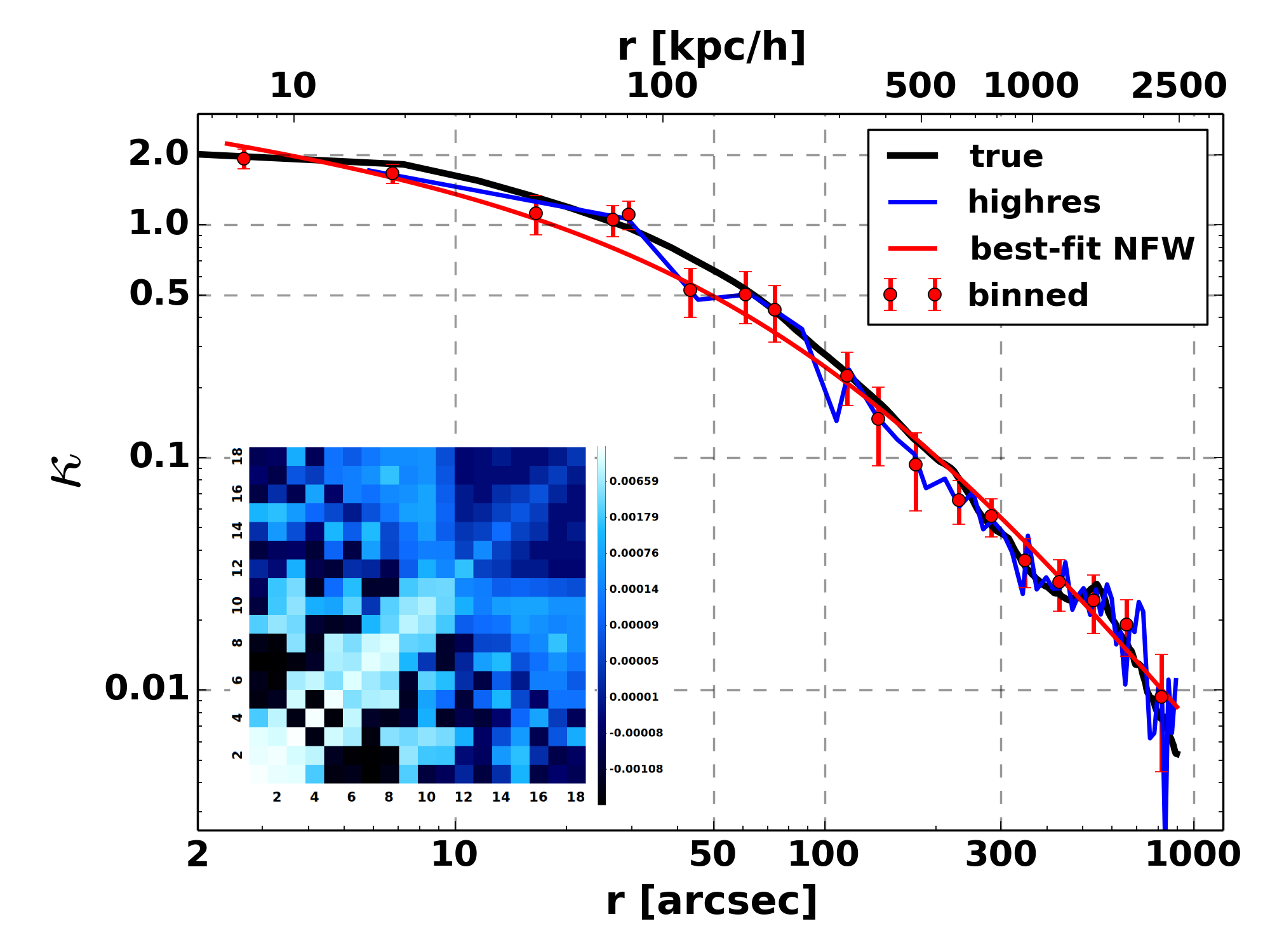}
     \caption{The convergence profile of the real cluster (black line) and of the reconstruction using two binning schemes for a source redshift of $z_{s}=2.0$. The fine binning scheme (blue line) follows nicely even smaller substructure components but is quite
      noise towards the outskirts. We use the logarithmic, coarser binning scheme (red points) to derive 2D NFW fitting parameters. The error bars are the square root of the diagonal elements in the bin-to-bin covariance matrix, which is shown in the bottom-left inset. The best-fit NFW profile is overplotted as the red line.}
     \label{fig_RAY_PROFILE}
    \end{figure}

    We derive surface-mass density profiles from both, the real projected matter distribution and our reconstruction. Shown in Fig.~\ref{fig_RAY_PROFILE} are the profiles as a function of distance from the densest point in the simulation.
     We perform two different radial binning schemes for our reconstruction. One fine, linear binning which is able to follow all substructure variations in the reconstruction, but gets very noisy
     especially in the outskirts of the field where the signal becomes weak. Hence, we also apply a coarser, logarithmically spaced binning scheme, which we will use later on for a parametric fit to the profile.
     The covariance matrix for the bins is derived from the profile analysis for 
     all of our 250 resampled realizations. 
     A first, visual,  inspection shows that our reconstructed profile follows nicely within the error bars the true profile.
      
     As a quantitative check we fit an NFW profile \citep{Navarro1996,Navarro1997} to the data. 
     Since we are not interested in the potential biases that the assumption of spherical symmetry introduces when fitting 2D data with profiles derived from 3D simulations \citep[see][and references therein for a discussion]{Meneghetti2014}, 
     we first derive the 2D NFW parameters for the true profile. We do so by assuming spherical symmetry while projecting the NFW profile and fitting it to the convergence profile. Following the commonly used
     parametrization of the NFW profile, this delivers a total mass $M_{200} = 1.90 \times 10^{15}M_{\odot}$ and a concentration $c_{200}= 5.46$ where both quantities are evaluated using a radius where the average density of the halo is 200 times the critical density of the Universe.
      By applying the same formalism to our reconstruction we find $M_{200} = 1.83 \pm 0.15 \times 10^{15}M_{\odot}$  and $c_{200}= 4.8  \pm 1.0$. Both, the reconstructed mass and concentration are in good agreement with the true 2D values within their $1\sigma$ error bars.      
     The best-fit, projected NFW profile is also overlaid in Fig.~\ref{fig_RAY_PROFILE}.
    
   \subsubsection{Magnification map}
\begin{figure}
      \includegraphics[width=.5\textwidth]{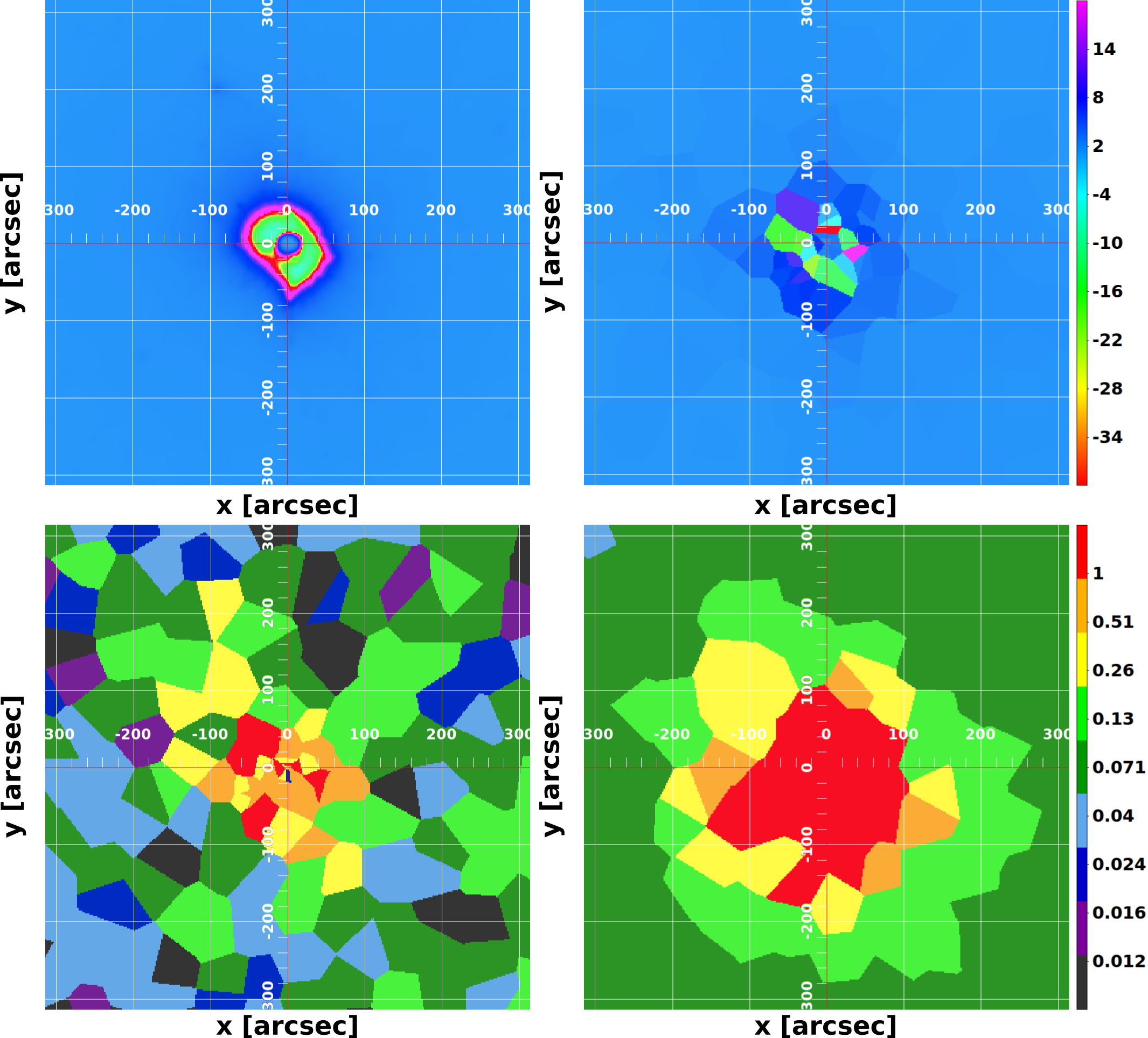}
     \caption{Real and reconstructed magnification maps in the \textit{top left} and \textit{top right panel}, respectively. All maps are scaled to a source redshift of $z_{\textrm{s}}=2.0$. The \textit{bottom left panel}
     shows the relative difference between the two maps above, evaluated at the positions of the reconstruction nodes. The \textit{bottom right panel} shows the standard deviation of the relative difference, as evaluated from the 250 reconstruction
     realizations.}
     \label{fig_RAY_magnification}
    \end{figure}

    For many applications including the exploration of the high-redshift Universe \citep[e.g.][]{Bradav2009,Zheng2012,Bradley2014} or the study of lensed supernovae \citep[e.g.][]{Patel2014,Kelly2015,Rodney2015a}, it is very important to have magnification estimates for a lens at specific locations in the image plane. In Fig.~\ref{fig_RAY_magnification}, 
      we show the real
      magnification map of the simulated halo as a zoom on the central region where the magnification starts deviating significantly from unity. 
      The magnification values are derived for a fixed source redshift of
      $z_{\textrm{s}}=2.0$. In the same figure we show the reconstructed magnification map for our fiducial reconstruction scaled to the same source redshift. For a quantitative comparison we show in the bottom panel of Fig.~\ref{fig_RAY_magnification} the relative error when comparing the reconstructed magnification value to the real one. The general
      trend of this comparison can be described as follows. In the outskirts of the field the difference to the true value is well below 10\% and mostly below 5\%. When approaching the innermost 200 arcsec of the field, the error increases to 10-20\%
      and increases further to about 50\% near and within the critical line of the cluster, which is at $\sim$50 arcsec at this redshift. For some reconstruction nodes the relative error can be well above 100\%. These large discrepancies are expected
      close to the critical line of the cluster, where the lensing effects become extremely non linear. To explore the significance of these deviations we calculate the relative errors in magnification for all of our reconstruction realizations and calculate
      the scatter in each reconstruction node. The result is also shown in Fig~\ref{fig_RAY_magnification}, indicating that all observed discrepancies between the reconstruction and the truth are well within the expected error bars.

     \subsubsection{Critical curve and Einstein radius}
    As the last analysis for the realistic lensing scenario we look into the reconstruction of the critical curve, again for a source redshift of $z_{\textrm{s}}=2.0$. Firstly, we calculate the true Einstein radius of the halo by sampling
     its critical line with 1040 points, calculating the distance of each sample point and taking the sample average. Following this approach the Einstein radius is $r_{E} = 46.5$ arcsec. For our fiducial reconstruction, we sequentially check for each
     pairing between a reconstruction node and its four nearest neighbors if the sign of the Jacobian determinant (Eq.~\ref{equ_jacobiandeterminant}) changes. If it does, we assign a point to the critical curve at the center of the connecting line between the node and the respective 
     neighbor.
     From this set of points we calculate the Einstein radius in the same way as described earlier. In order to assign an error bar, we repeat the procedure for all reconstruction realizations. The resulting
     Einstein radius of $r_{E} = 47 \pm 4$ arcsec is in excellent agreement with the real value.
      
\begin{figure}
      \includegraphics[width=.5\textwidth]{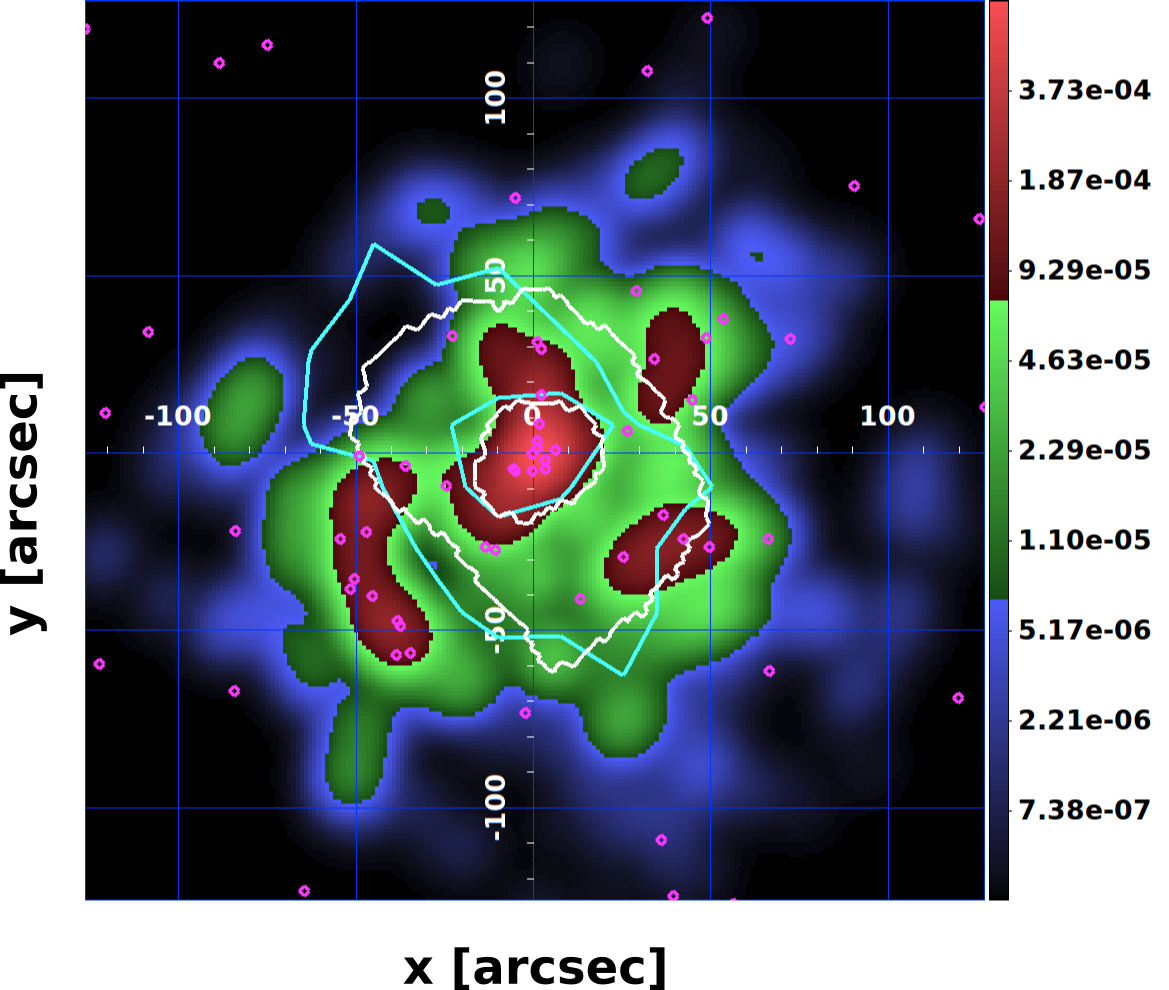}
     \caption{The critical curves of the halo for a source redshift of $z_{\textrm{s}}=2.0$. The white line shows the true position, while the cyan line shows the reconstruction from our fiducial model.
      The color of the background encodes the probability distribution function of a point in the lens plane to be part of the critical curve in the reconstruction. The pink circles show the distribution of reconstruction nodes near the critical
curve of the halo.}
     \label{fig_RAY_ccurve}
    \end{figure}

     We show the full critical curve of the system in Fig.~\ref{fig_RAY_ccurve}, together with the reconstruction from the fiducial model.
     The color-coded background of the figure shows the probability distribution function (PDF) for a point in the lens plane to be part of the critical curve. 
     It is derived from a Gaussian kernel density estimator applied to the distribution of points assigned as part of the critical curve for all 250 reconstruction realizations. 
     One can see that the areas of highest probability nicely follow the real position of the critical curve with two exceptions. The upper-left edge of the critical line, around coordinate (-50,50) arcsec is
     not recovered at all and there is a clear misidentification of the critical line around the coordinate at (-50,-50) arcsec. We think that these two discrepancies can be explained with the distribution of the reconstruction nodes, which is also
     overlaid in the figure. The part of the critical curve which is recovered by the fiducial model, but seems unlikely given the error model clearly lies in an area of the field which is not well sampled by the reconstruction nodes.
     The opposite is the case where the contours of high probability clearly deviate from the real critical curve. Here the dense sampling of reconstruction nodes seems to steer some reconstruction realizations away from the true position. 
     Although the general recovery of the critical curve, especially for the fiducial model, is good, this result motivates the exploration of alternative reconstruction node placements, which 
     we reserve for a more detailed follow up work.
    
    This finalizes our collection of realistic tests to determine the performance of the new, RBF-based methodology. We have shown that the method performs well in recovering the density profile, NFW parameters, the magnification map, the Einstein radius and with some caveats
      also the critical curve of the simulation. However, we want to point out that future studies are necessary beyond the general presentation of methodology in this work. Upcoming tests will include a much less massive lens, bigger variations
      in the number of multiple images and the effect of catastrophic redshift errors. Furthermore, we want to investigate alternative strategies of placing the reconstruction nodes, especially in the strong lensing regime.
    
   \section{Conclusions}
   \label{sec_conclusions} 
   
   In this work we introduced a framework for the mesh-free interpolation and differentiation of functions. 
   This framework is an important and potentially powerful tool for applications
   in the field of gravitational lensing, because input data does usually not follow any regular pattern in its spatial distribution
   and is confined to very different length scales. 
   The particular examples for this problem in this work are the regimes of weak and strong gravitational lensing.
   
   Our implementation of mesh-free interpolation and differentiation is based on the concept of radial basis functions (RBFs). 
   Specifically, we use Gaussian radial basis functions, although our methodology is not restricted to this one class
   of RBFs. We convincingly proved the performance of our implementation in Secs.~\ref{sec_rbf} and Apps.~\ref{app_rbf_inter} and
   \ref{app_rbf_fd}. We showed the importance of a well vetted shape parameter of the Gaussian RBF, depending on evaluation-points layout 
   and application and showed that by using an increasing number  of nodes in the nearest-neighbors stencils, higher
   accuracies can be achieved with the drawback of longer runtime. If all these parameters are chosen appropriately, the accuracy
   of our interpolation and differentiation routines is well below the percent level.
   
   Using the new techniques to express mesh-free numerical derivatives we implemented a novel method for mass
   reconstruction from gravitational lensing. We translated the initial ideas of \citet{Bartelmann1996, Bradav2005, Cacciato2006}
   and \citet{Merten2009} into the realm of a mesh-free and intrinsically adaptive reconstruction and
   tested the performance of this approach with a simple mock lens. 
   In Sec.~\ref{sec_testing_toy_model} we showed with a simple toy-model lens that we are able to reconstruct the lens based on 55 multiple images
   Weak-lensing constraints, when used as a stand-alone input, allow a reconstruction also at the $5$--$10\%$ accuracy level but do not cover the strong lensing core of the lens. 
   The main problem in weak lensing analyses though is the presence of shape noise in the input data. We showed the performance
   of the well-established two-level iteration scheme of \citet{Bradav2005} and \citet{Merten2009} in order to deal with 
   noisy data in Fig.~\ref{fig_WL_rec_Noise_iter} and showed our ability to account for the redshift distribution of sources in Fig.~\ref{fig_WL_rec_z_corr}. Finally, we combine
   weak -and strong lensing constraints to achieve an accurate reconstruction of the mock lens over all relevant length scales. A much more relevant 
   test in terms of applicability to real data was performed in Sec.~\ref{sec_ray}. We showed that we can reproduce the mass of a simulated, massive halo with a precision
   of 8\% while using realistic weak and strong lensing input constraints. The same precision is achieved in the recovery of the Einstein radius. While we also recover the
   concentration, magnification map and the critical curve of the halo within our error bars, the results are less precise. Future studies will show how they will improve with 
   a larger number of strong lensing input constraints. An important test, for example in the context of the on-going HST Frontier Fields\footnote{\href{http://www.stsci.edu/hst/campaigns/frontier-fields/}{http://www.stsci.edu/hst/campaigns/frontier-fields/}}.

   An application to known, real observed lenses has to follow together with a comparison to other reconstruction techniques based on gravitational lensing.
   Furthermore, we are planning to apply our method not only to mass-reconstruction
   applications but extend our work to lens-source plane mapping. While lensing features are usually observed on very regular meshes
   in the lens plane, due to the pixelization scheme of CCD images, the lens mapping transforms this regular pattern
   into a very irregular one in the source plane. This is a problem both in ray-tracing simulations \citep{Meneghetti2008}
   and in the source-plane reconstructions of lensed sources \citep[see e.g.][and references therein]{Dye2005,Vegetti2009,Tagore2014}
   Finally, we will apply our reconstruction method to real data and we will explore the usefulness of our approach in the
   field of PSF interpolation, which also deals with the irregular pattern of star positions in the fields of astronomical observations.
   Improvements to our general implementation may stem from the introduction of a spatially varying shape parameter \citep{Fornberg2007},
   although most recent developments in applied mathematics may allows us to discard the shape parameter altogether when using 
   polyharmonic spline-type RBFs (Bengt Fornberg, private communication).

\section*{Acknowledgements}
I want to send a warm thank you to Bengt Fornberg for helping me to understand and 
implement the concept of radial basis functions for the purpose of finite differencing.
I also thank Matthias Bartelmann, Massimo Meneghetti, 
and Leonidas Moustakas for inspiring discussions.
This research was carried out at the Jet Propulsion Laboratory, 
California Institute of Technology, under a contract with NASA and 
I acknowledge support from NASA Grants HST-GO-13343.05-A
and HST-GO-13386.13-A. The research
leading to these results has received funding
from the People Programme (Marie Curie Actions) of
the European Union’s Seventh Framework Programme
(FP7/2007-2013) under REA grant agreement number
627288.

%\bibliographystyle{mnras}
%\bibliography{mertenbibs.bib}  
 
 \input{rbf_1.bbl}

  \appendix
  \section{Interpolation with radial basis functions}
  \label{app_rbf_inter}
  In the following we test the robustness and accuracy of the interpolation in 2D using a Gaussian radial basis function (RBF). 
  The main free parameters in this analysis are the shape parameter of the Gaussian RBF, the number of nearest neighbors
  used in the interpolation stencil and the number of evaluation points to perform the interpolation. 
 
\begin{figure}
    
    \centering
      \includegraphics[width=.475\textwidth]{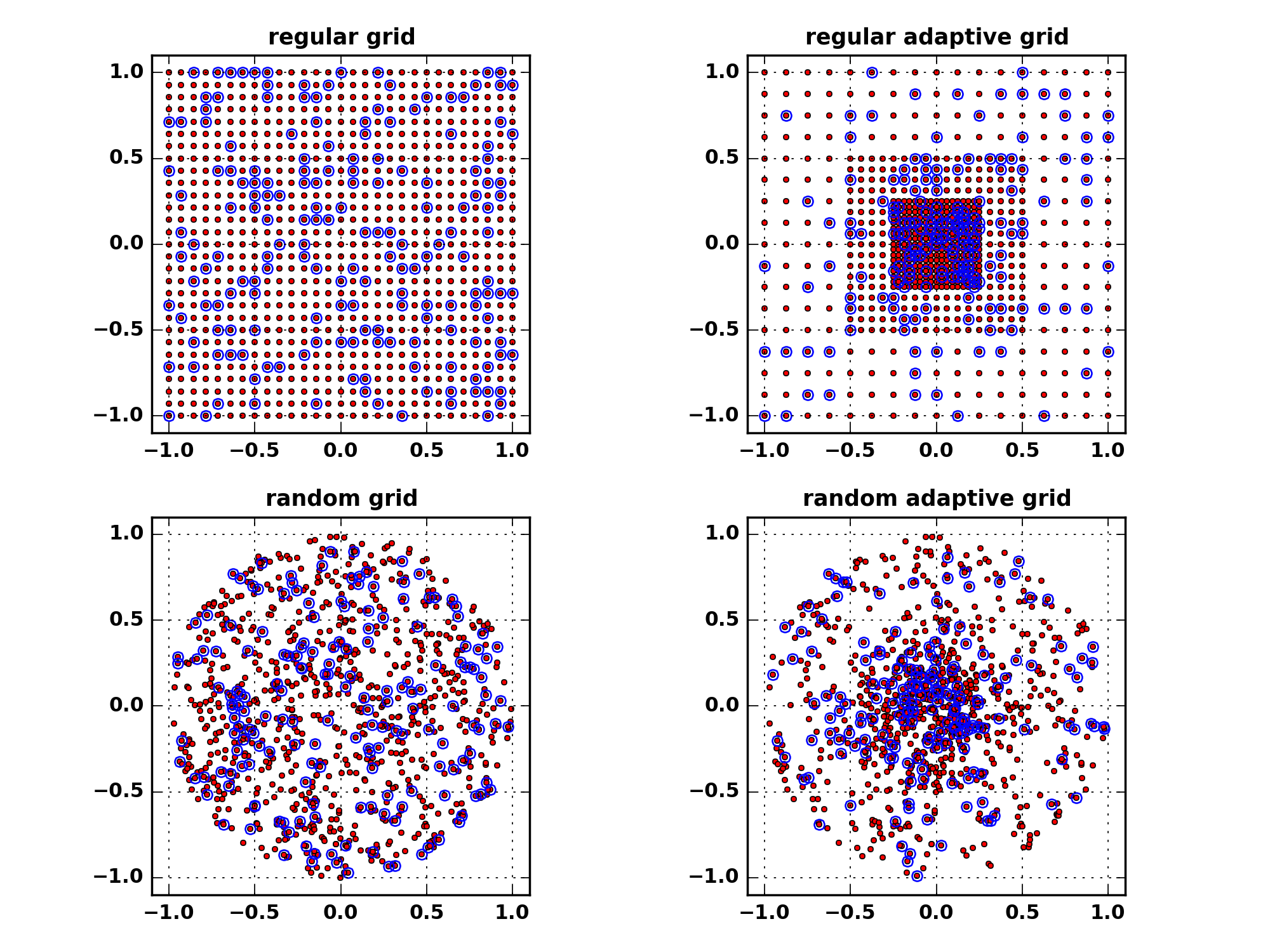}
     \caption{The evaluation point configurations we use for our performance tests of RBF interpolation and differentiation. The red points
     in the top left panel show a regular mesh with 900 nodes and no refinement. The top right panel shows a regular mesh
     of 900 nodes with two refinement levels towards the center. The bottom panels show 900 randomly chosen nodes on the unit disk. 
     The example in the bottom left panel is not refined, while the bottom right panel includes two levels
     of refinement towards the center of the disk. In each refinement step, the density of random points doubles. The blue 
     circles show an ensemble of 180 evaluation points for each setup, which anchor the interpolant in the interpolation
     test.}
     \label{fig_appendix_grids}
    \end{figure}

\begin{figure}
    \centering
      \includegraphics[width=.5\textwidth]{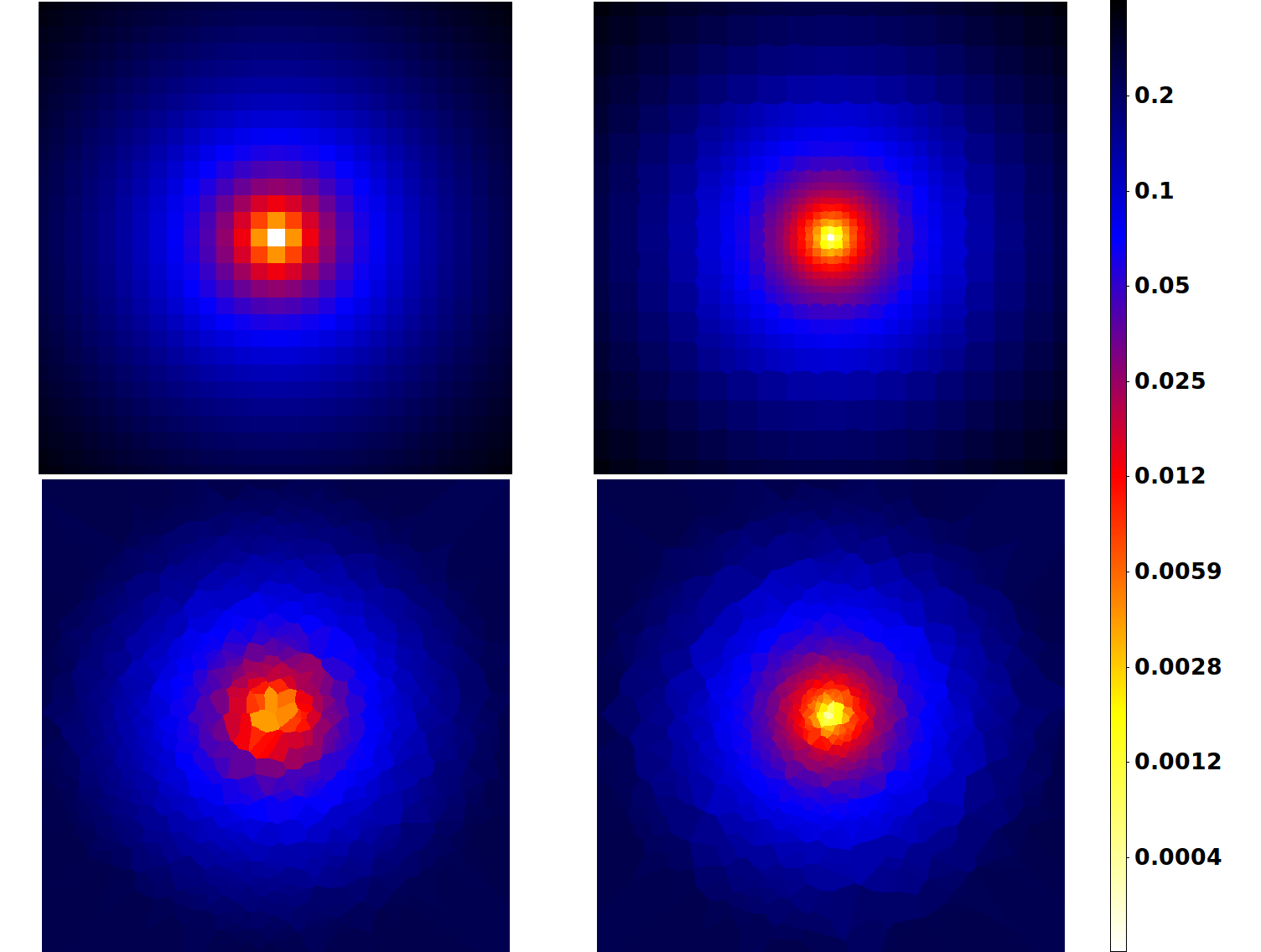}
     \caption{The test function defined in Eq.~\ref{nfw_test_function} evaluated at the nodes of the four setups shown in 
      Fig.~\ref{fig_appendix_grids}.}
     \label{fig_appendix_grids_function}
    \end{figure}

 We define the 2D lensing potential of an NFW halo \citep[e.g.][]{Bartelmann1996,Meneghetti2000,Wright2000}  $\psi(x,y) = 4\kappa_{\textrm{s}}a(x,y)$ as a test function, with
 \begin{equation} \label{nfw_test_function}
  a(x,y) = \frac{1}{2}\ln^{2}\frac{r}{2} + \left\{ \begin{aligned}
2\arctan^{2}\sqrt{\frac{r-1}{r+1}}&\quad \textrm{for}&(r>1)\\
-2\arctanh^{2}\sqrt{\frac{1-r}{1+r}}&\quad \textrm{for}&(r<1)\\
0&\quad\textrm{for}&(r=1)\\
\end{aligned}
\right.
 \end{equation}
and $r=\sqrt{x^{2}+y^{2}}$. For the scale convergence we choose $\kappa_{s}=0.25$
  and define four different kind of node layouts, all of which are shown in the four panels of Fig.~\ref{fig_appendix_grids}. 
  The first layout is a regular, square mesh with 900 nodes. The second mesh has the same number of total nodes but is refined
  twice towards the center of the mesh. In each refinement step, the separation of neighboring mesh nodes is reduced by a factor 
  of two. The third layout is defined by 900 random nodes on the unit disk, as is the fourth layout, with the difference that also this
  configuration is refined twice towards the center of the disk. The last two examples define mesh-free sets of nodes.  
  In the following, we interpolate the test function
  on these domains. For illustration purposes we evaluated Eq.~\ref{nfw_test_function} on the nodes of all four configurations in 
  Fig.~\ref{fig_appendix_grids_function}.
  
\begin{figure}
    \centering
      \includegraphics[width=.475\textwidth]{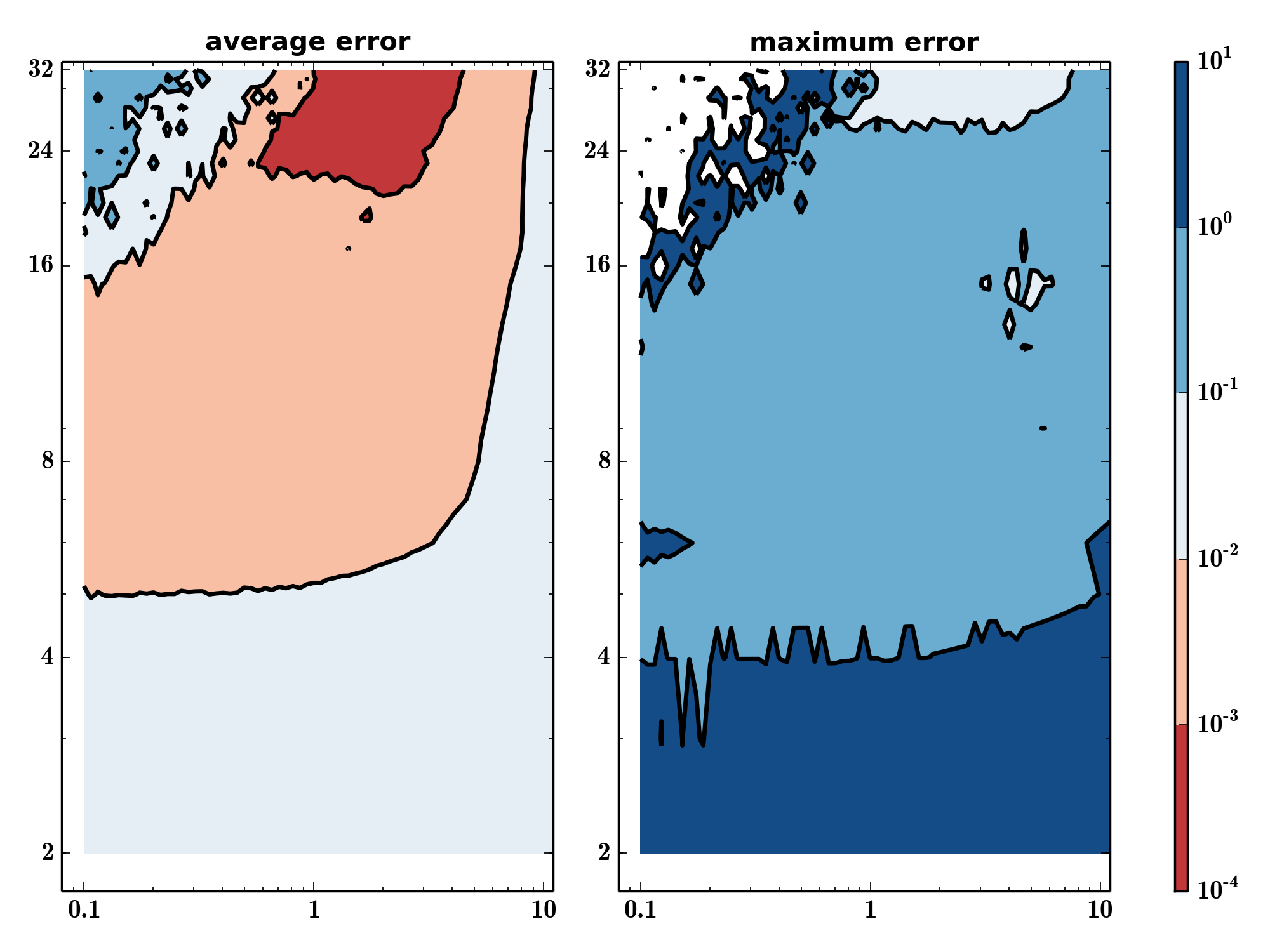}
     \caption{Interpolation performance on a regular, unrefined mesh with 180 evaluation points. The left panel shows the average relative
      interpolation error as a function of the shape parameter of the Gaussian RBF and the number of nearest neighbors used
      to calculate the interpolant in each grid point. The right panel shows the maximum relative error of all nodes.}
     \label{fig_appendix_inter_reg}
    \end{figure}

\begin{figure}
    \centering
      \includegraphics[width=.475\textwidth]{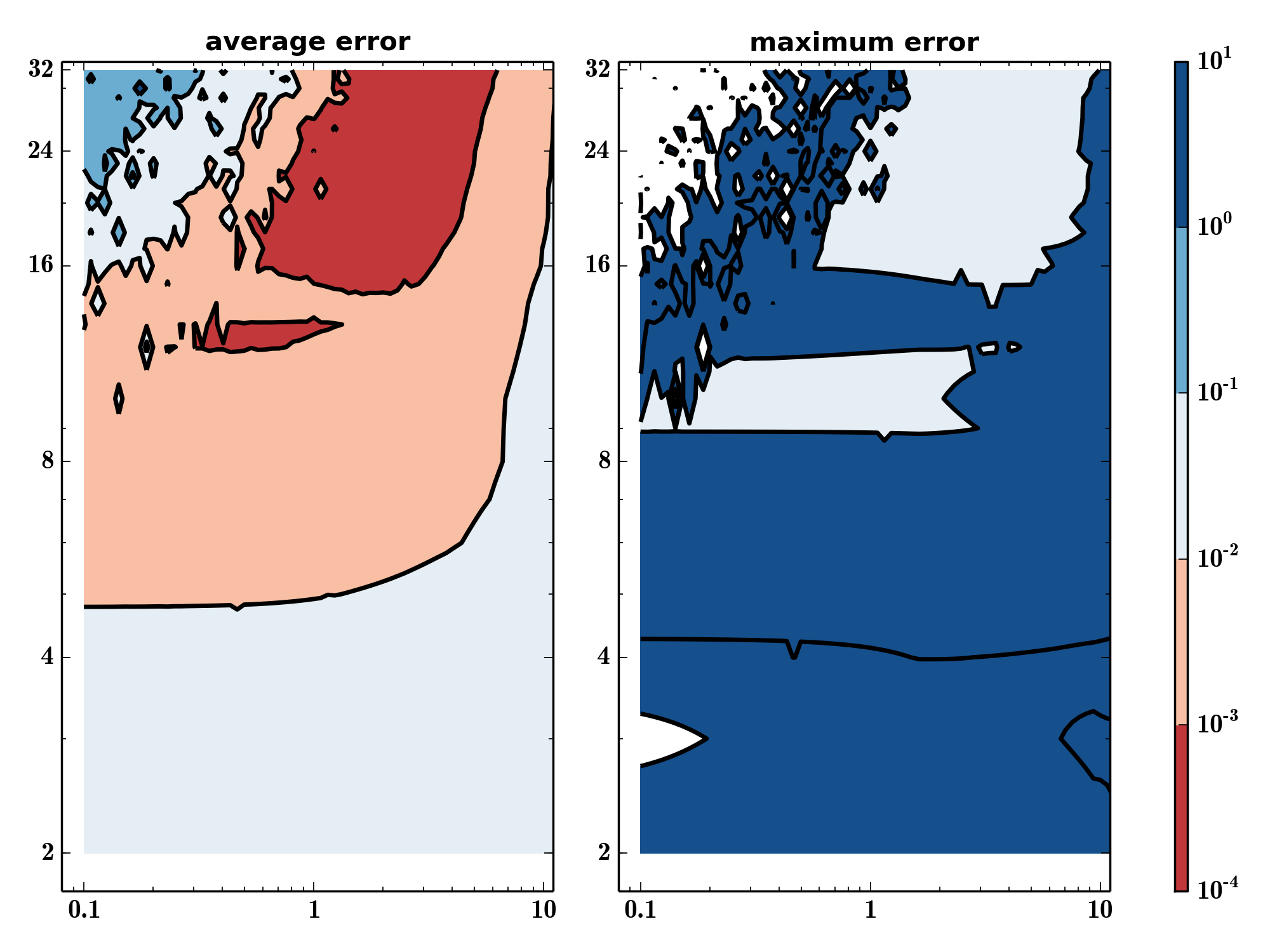}
     \caption{This figure shows the same plot as in Fig.~\ref{fig_appendix_inter_reg}, but for a mesh-free, unrefined node layout with 
      180 evaluation points.} 
     \label{fig_appendix_inter_ran}
    \end{figure}

\begin{figure}
    \centering
      \includegraphics[width=.475\textwidth]{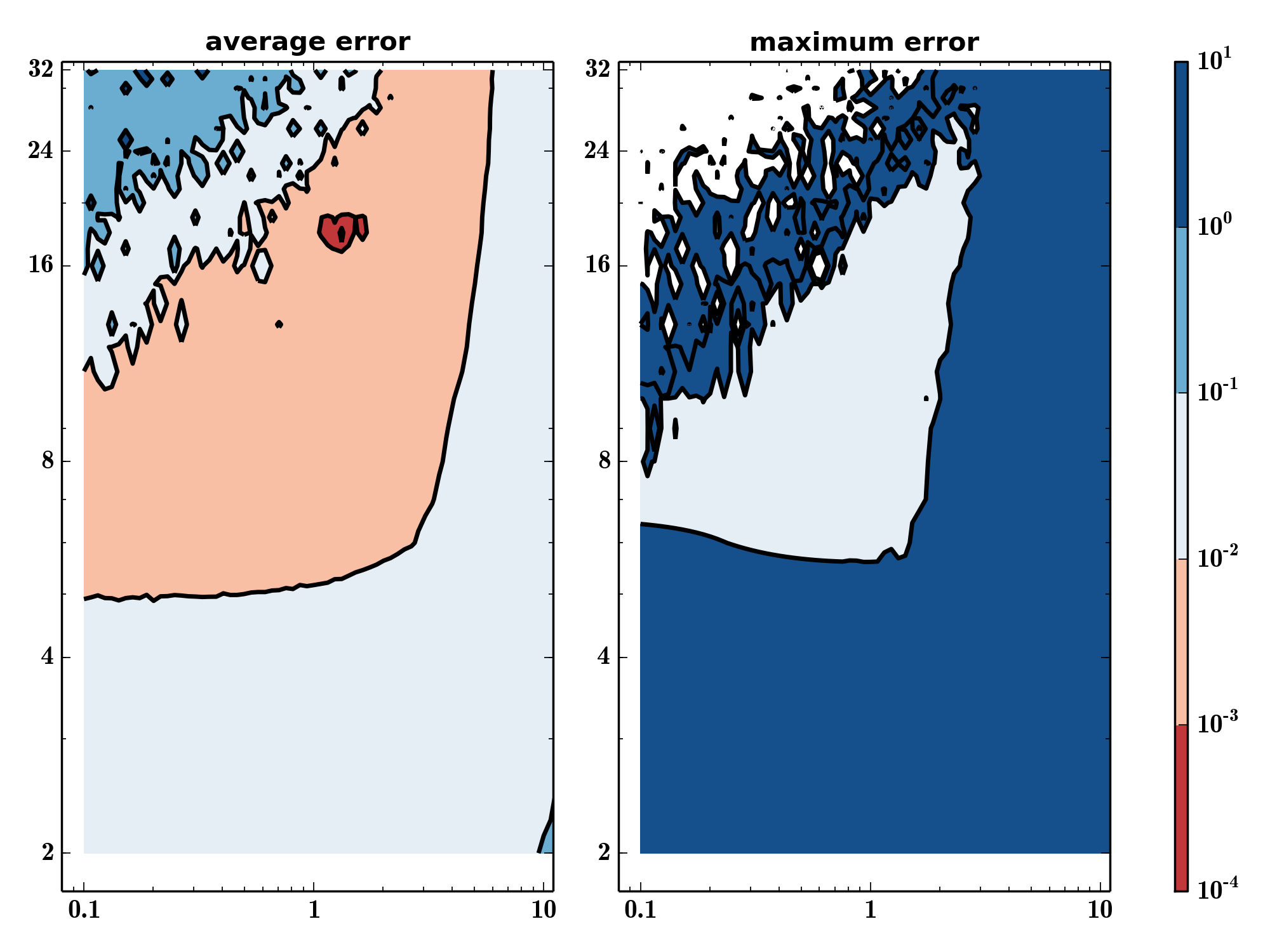}
      \caption{This figure shows the same plot as in Fig.~\ref{fig_appendix_inter_reg}, but for a refined
      regular mesh with 180 evaluation points.}
      \label{fig_appendix_inter_reg_adapt}
    \end{figure}

\begin{figure}
    \centering
      \includegraphics[width=.475\textwidth]{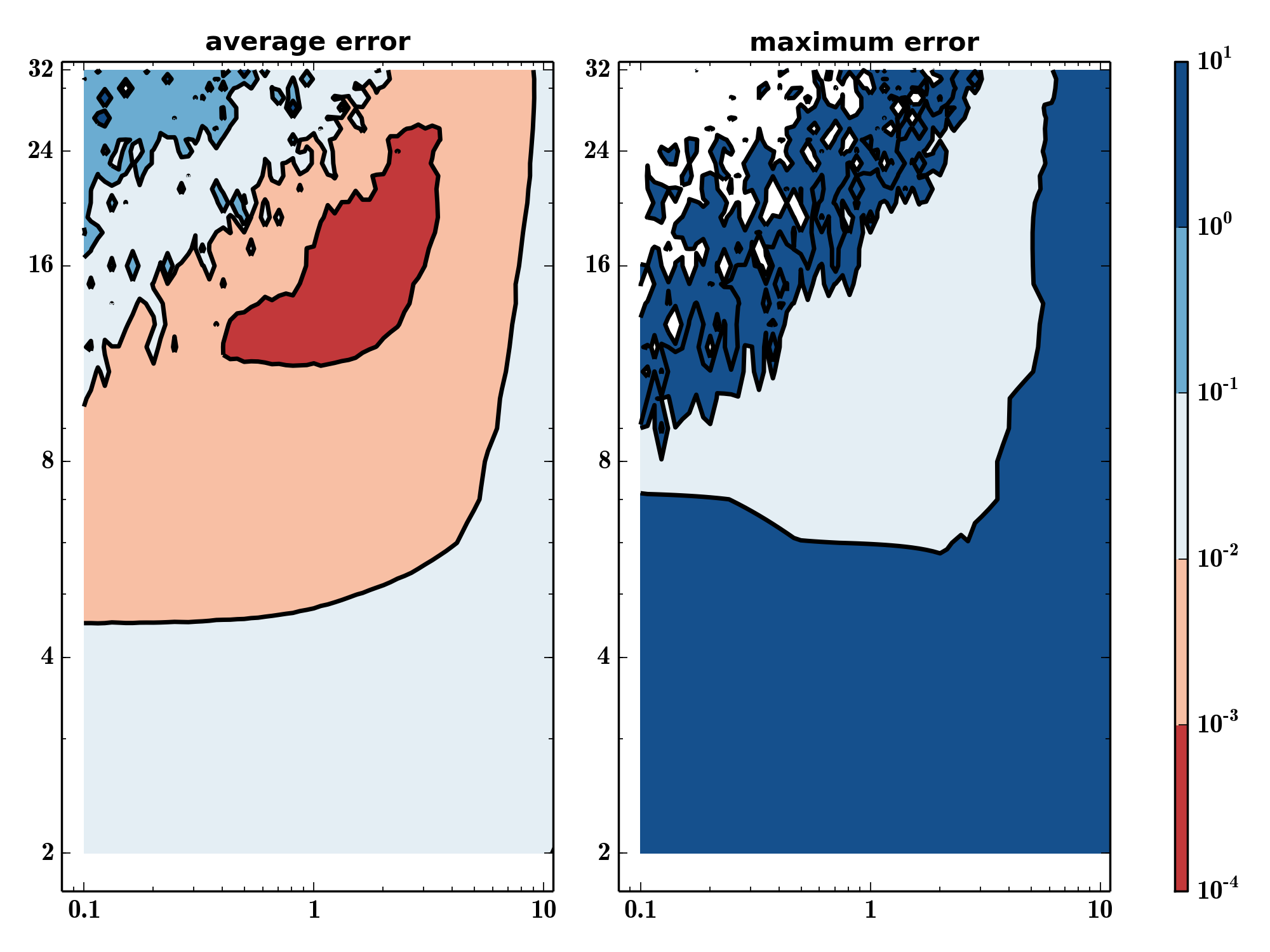}
      \caption{This figure shows the same plot as in Fig.~\ref{fig_appendix_inter_reg}, but for a random, mesh-free node layout 
      with two levels of refinement towards the center.}
      \label{fig_appendix_inter_ran_adapt}
     \end{figure}

  In order to perform interpolations, we define as a first step a set of 180 evaluation points in each of the four test cases. 
  The evaluation points for each node layout are shown as blue circles in Fig.~\ref{fig_appendix_grids}.
  We calculate the interpolant $\tilde f(x,y)$ of the test function $f(x,y)$ by using Eq.~\ref{equ_rbf_interpol} and define
  two performance metrics. The average relative error $\left\langle(\tilde f(x,y) - f(x,y))/f(x,y)\right\rangle$ for all nodes
  $\vec{x}=(x,y)$ and the maximum relative error $\max((\tilde f(x,y) - f(x,y))/f(x,y))$ for 
  all nodes $\vec{x}=(x,y)$. We evaluate both metrics as a function of the shape parameter and the number
  of nearest neighbors used to derive the interpolant at a given node $\vec{x}$. For the regular, nonrefined mesh we plot
  the results in Fig.~\ref{fig_appendix_inter_reg} and for the nonrefined, mesh-free setup in Fig.~\ref{fig_appendix_inter_ran}. 
  As one can see, the overall performance is excellent once the right shape parameter is found. With the use
  of 16 nearest neighbors or more, the average relative errors approach the $10^{-4}$ level and the maximum
  error approaches $10^{-2}$. The performance is slightly better for the mesh-free setup, which is due to the fact that the 
  RBF is approach is not well suited to treat the edges of the regular mesh. The same holds for the interpolation on the
  refined node layouts. We show the results for the regular, refined mesh in Fig.~\ref{fig_appendix_inter_reg_adapt} and for the random,
  mesh-free node layout in Fig.~\ref{fig_appendix_inter_ran_adapt}. The performance in the latter case is similar to the unrefined one,
  but the performance drops slightly for the regular, refined mesh. Also here the RBF approach is not ideal to treat the abrupt transitions
  between the different refinement levels, which are not well-described by a radially dependent function. However, the 
  overall performance in all four cases is remarkable. 
  
\begin{figure}
      \includegraphics[width=.475\textwidth]{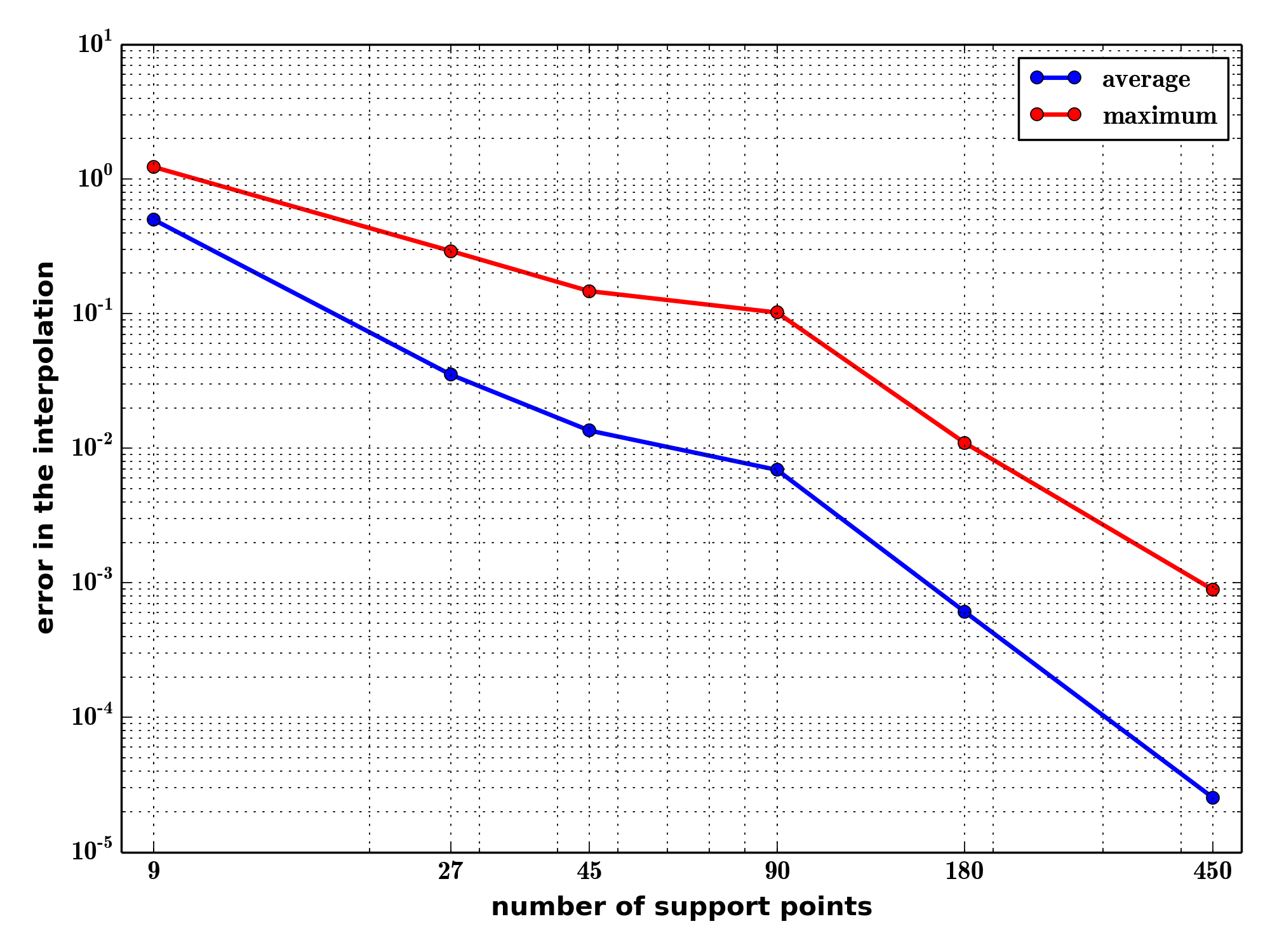}
     \caption{The accuracy of the interpolation on the mesh-free, refined node layout as a function of the number of evaluation points. 
      The shape parameter and the number of nearest neighbors were chose to be optimal, according to the previous analysis.}
     \label{fig_appendix_inter_support}
    \end{figure}

\begin{figure}
    \centering
      \includegraphics[width=.475\textwidth]{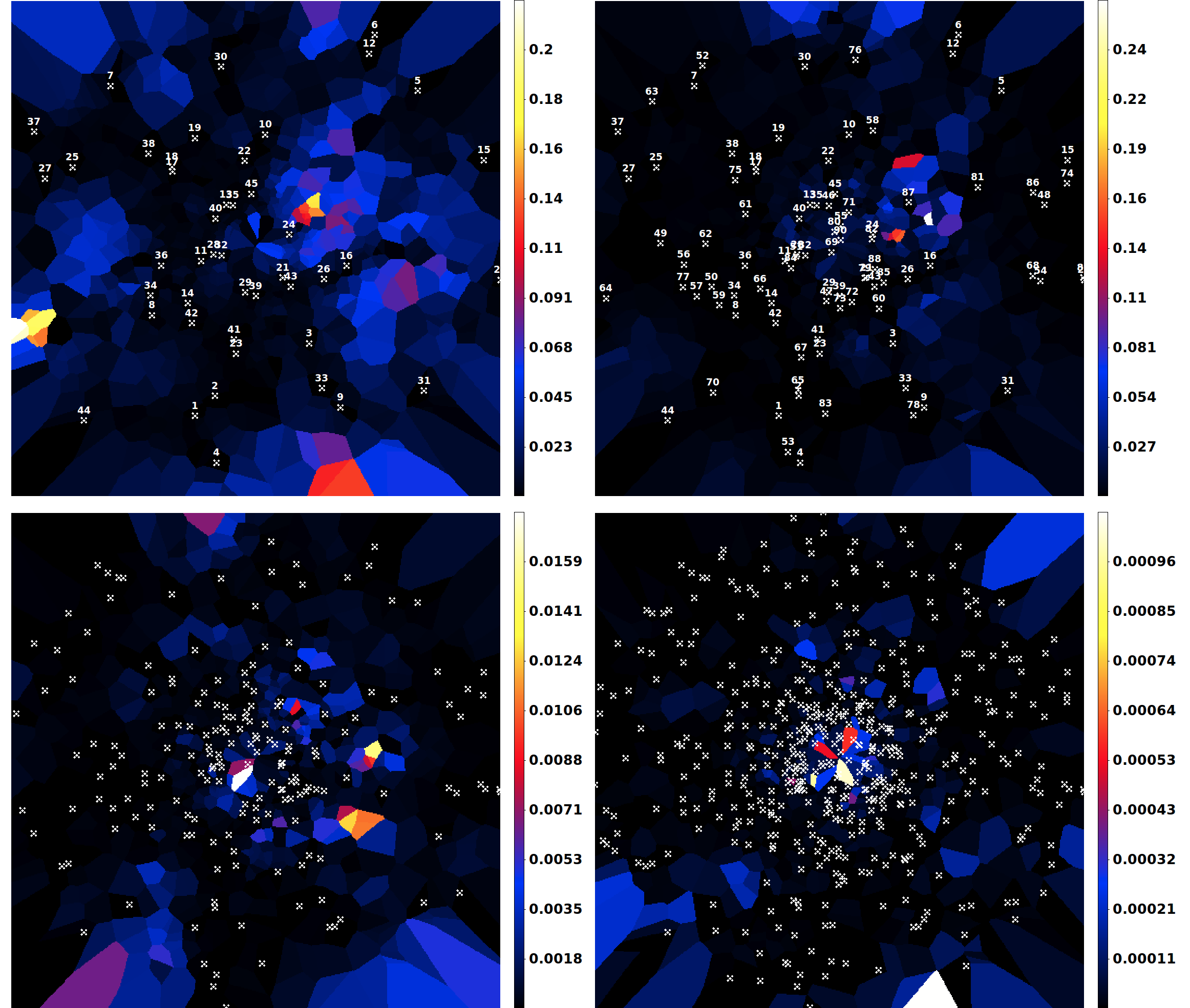}
     \caption{A visualization of the interpolation performance for 45, 90, 180 and 450 evaluation points, respectively. 
     The color coding shows the relative interpolation error at each points of the mesh-free, refined node layout. 
     The white markers in each panel show the the position of the evaluation points.}
      \label{fig_appendix_inter_pos}	
    \end{figure}

  As a last test we vary the number of evaluation points in the mesh-free, refined node layout, going from 1\% of the total number of nodes, 9, 
  to 50\% of the total number of nodes, 450. We plot both performance metrics as a function of the number of evaluation points in 
  Fig.~\ref{fig_appendix_inter_support}, where the shape parameters and the number of nearest neighbors in each case 
  are optimally chosen. The interpolation accuracy is clearly a steep function of the number
  of evaluation points, where very good results ($<10^{-4}$ average and $<10^{-2}$ maximum relative error) can be achieved
  with a large number of evaluation points exceeding one quarter of the total number of nodes.  We investigate the dependence of the interpolation
  accuracy on the number and position of the support points a little further in Fig.~\ref{fig_appendix_inter_pos}, where
  we show the relative interpolation error for four different evaluation point setups.

  \section{Finite differencing with radial basis functions}
   \label{app_rbf_fd}
   
   To test the accuracy of numerical differentiation using radial basis function finite differencing stencils (RBF FD), we return
   to Eq.~\ref{nfw_test_function} and calculate two important derivatives.
   \begin{equation}\label{nfw_test_function_dx}
    \frac{\partial \psi(x,y)}{\partial x} = \alpha_{1}(x,y) = x\frac{\kappa_{\textrm{s}}}{r^{2}}b(x,y)
   \end{equation}
   with
    \begin{equation}
  b(x,y) = \frac{1}{2}\ln\frac{x}{2} + \left\{ \begin{aligned}
\frac{2}{\sqrt{r^{2}-1}}\arctan\sqrt{\frac{x-1}{x+1}}&\quad \textrm{for}&(x>1)\\
\frac{2}{\sqrt{1-r^{2}}}\arctanh\sqrt{\frac{1-x}{1+x}}&\quad \textrm{for}&(x<1)\\
1&\quad\textrm{for}&(x=1)\\
\end{aligned}
\right.
 \end{equation}\label{nfw_test_function_laplace}
   which is the first component of the deflection angle, and
  \begin{equation}
   \frac{1}{2}\bigtriangleup \psi(x,y) = \kappa(x,y) = 2\frac{\kappa_{\textrm{s}}}{r^{2}-1}c(x,y)
  \end{equation}
  with
   \begin{equation}
  c(x,y) = \left\{ \begin{aligned}
1-\frac{2}{\sqrt{r^{2}-1}}\arctan\sqrt{\frac{x-1}{x+1}}&\quad \textrm{for}&(x>1)\\
1-\frac{2}{\sqrt{1-r^{2}}}\arctanh\sqrt{\frac{1-x}{1+x}}&\quad \textrm{for}&(x<1)\\
\frac{1}{3}&\quad\textrm{for}&(x=1)\\
\end{aligned}
\right.
 \end{equation}
which is the convergence of the NFW test potential.

\begin{figure}
    \centering
      \includegraphics[width=.475\textwidth]{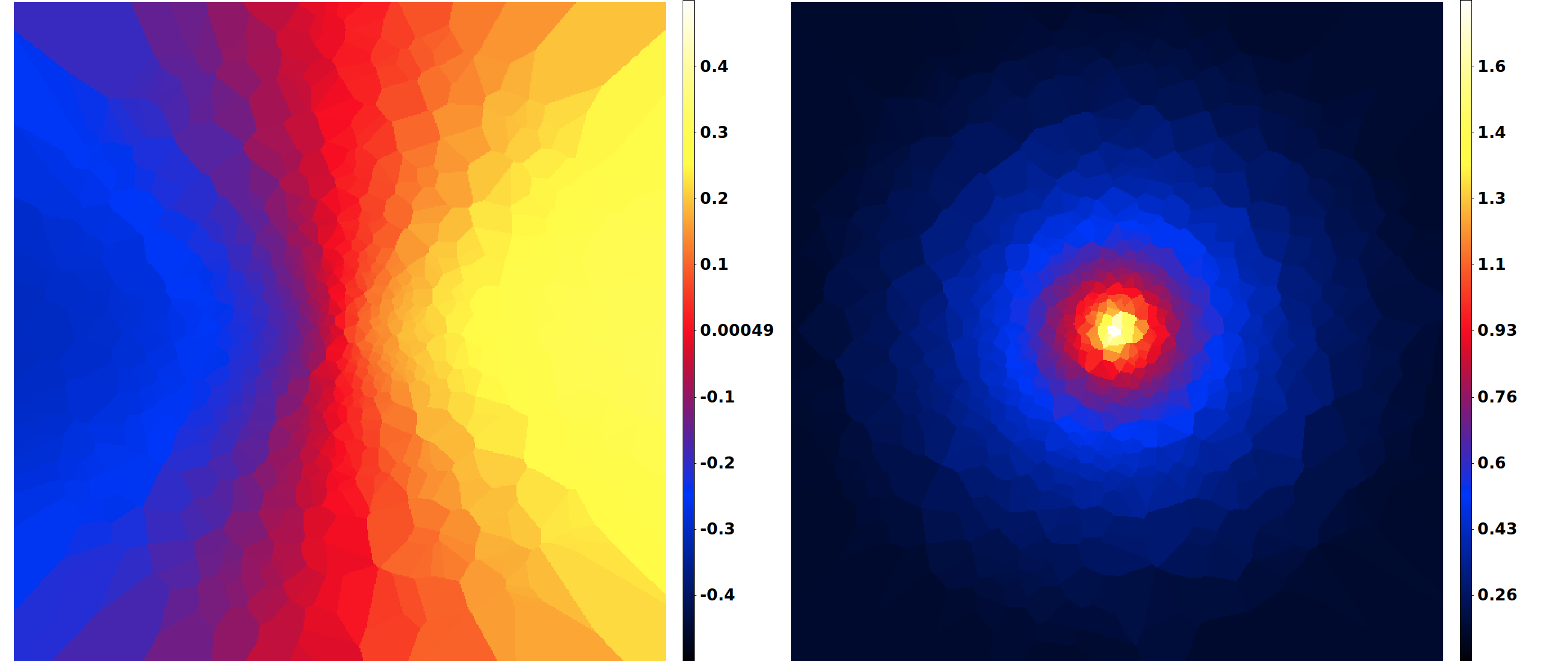}
     \caption{The derivatives of the test function on the mesh-free, refined node layout.  
     The left panel shows the first derivative in the x-direction and the right panel shows the 2D Laplacian
     of the function multiplied by 1/2.}
     \label{fig_appendix_fd_function}
    \end{figure}

   In the following we perform tests on the mesh-free refined node layout of App.~\ref{app_rbf_inter} only, since it resembles best
   the conditions in our real lensing applications. We visualize the test function and the three derivatives of interest for this node configuration
   in Fig.~\ref{fig_appendix_fd_function}.	 
  
\begin{figure}
    \centering
      \includegraphics[width=.475\textwidth]{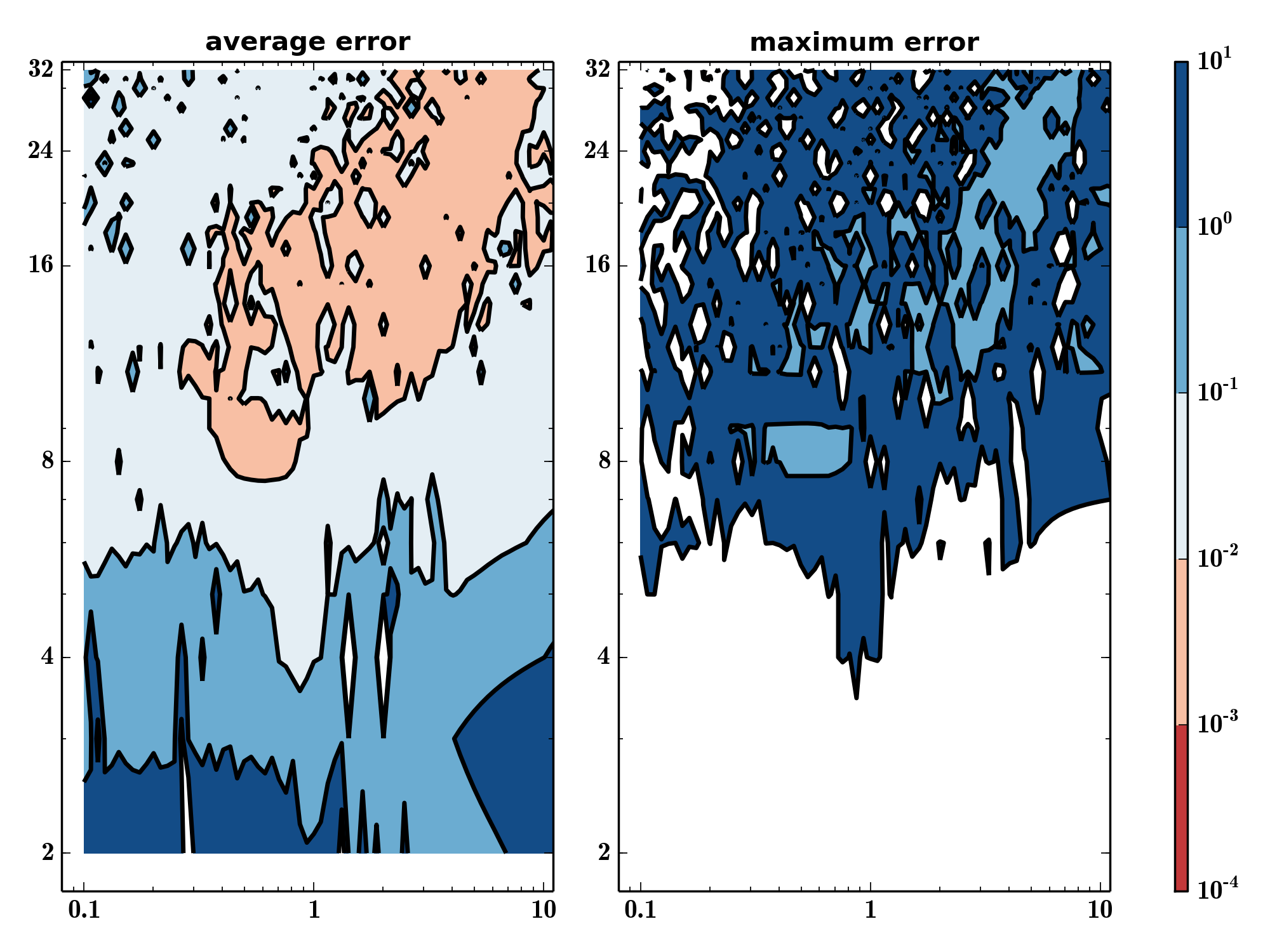}
     \caption{The accuracy of the numerical derivatives calculated by RBF FD. Shown is the average and maximum relative 
     error on the mesh-free, refined node layout for the first x derivative of the test function.}
      \label{fig_appendix_fd_dx}	
    \end{figure}

\begin{figure}
    \centering
      \includegraphics[width=.475\textwidth]{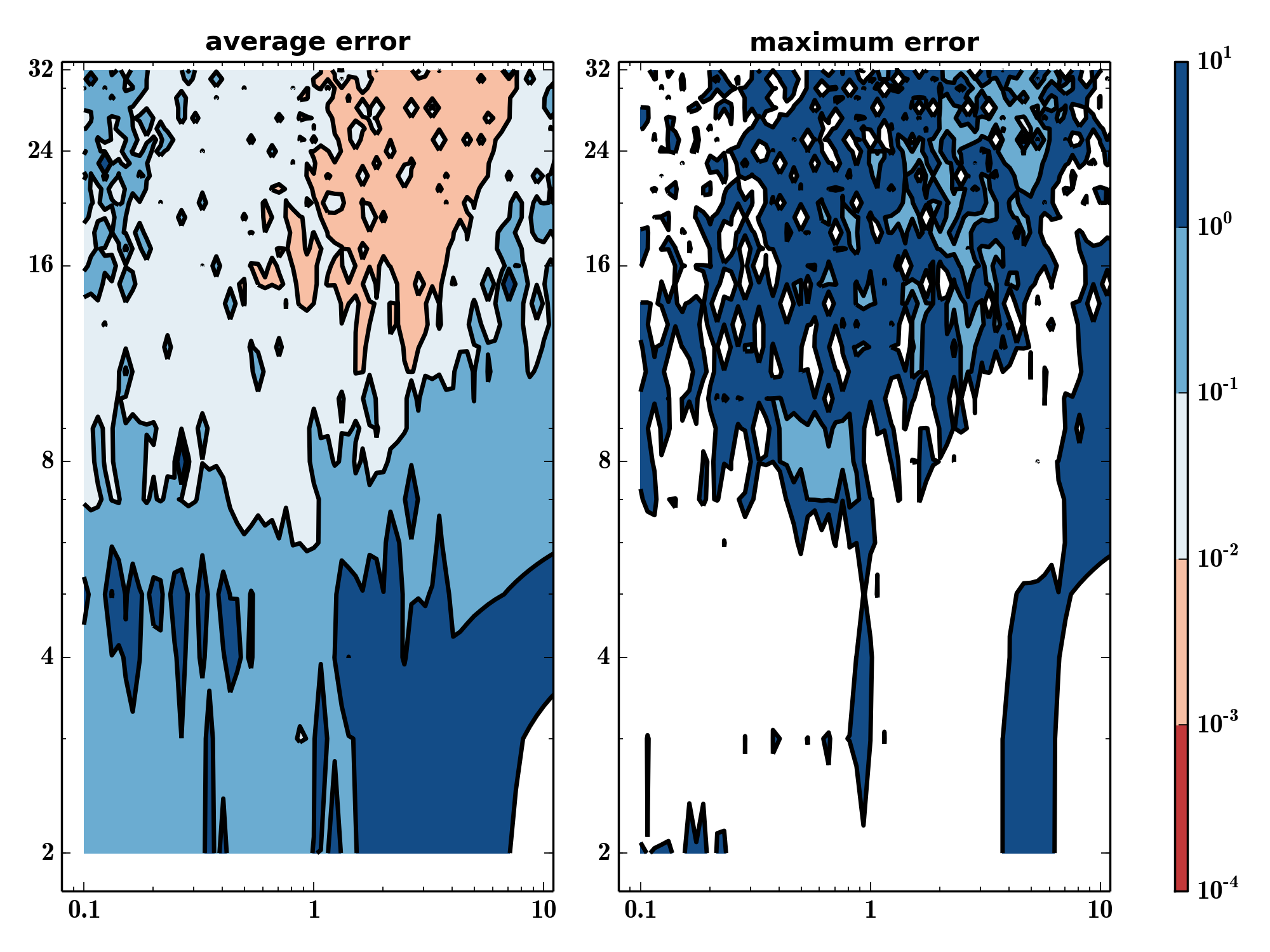}
     \caption{This figure is identical to Fig.~\ref{fig_appendix_fd_dx} but shows the RBF FD performance for the 2D Laplacian
     of the test function multiplied by 1/2.}
     \label{fig_appendix_fd_laplace}	
    \end{figure}

    We use Eq.~\ref{equ_FD} to calculate the derivatives of the test function. This operation
    has the Gaussian RBF shape parameter and the number of nearest neighbors as free parameters.
    We again define the average and the maximum error for all nodes as performance metrics and plot them for the first x derivative
    in Fig.~\ref{fig_appendix_fd_dx}. As expected, the right choice of the shape parameter is crucial and the accuracy steadily
    increases when using more nearest neighbors to calculate the numerical derivatives. This also applies to the accuracy of the
    numerical Laplacian as shown in Fig.~\ref{fig_appendix_fd_laplace}. With the right choice of
    shape parameter and an adequate number of nearest neighbors, average relative errors on the grid of $<10^{-3}$ and maximum
    errors of $<10^{-1}$ can be achieved throughout.
    
    \section{Linearization of the likelihood function}
    \label{app_lse}
    In the methodology outlined in Sec.~\ref{subsec_unstrucgrid_lensing} we need to minimize a complicated $\chi^{2}$-function 
    with the lensing potential in each evaluation point as a free parameter. This function 
    consists of a contribution from a  reduced-shear term, a critical-line estimator
    term, a multiple-image system term and a regularization term. The solution to this system is the 
    wanted lensing potential. Many of the explicit
    calculations have already been carried out in \citet{Bradav2005} and \citet{Merten2009} which is why we only carry out 
    the derivation of the multiple-image system term, not part of \citet{Merten2009} and where we use a slightly different approach than \citet{Bradav2005}. 

    Starting point is Eq.~\ref{equ_chi2MSYS} which we partially differentiate after $\psi_{k}$ and use Eq.~\ref{equ_pot_alpha} and Eq.~\ref{equ_pot_delta}.
\begin{equation}
 \begin{split}
   \frac{\partial\chi^{2}_{\mathsf m}}{\partial\psi_{l}} = \sum\limits_{n=1}^{N_{\mathrm{s}}}\frac{2}{\sigma^{2}}&\left[\theta_{n}\mathcal{D}_{nl}+\mathcal{D}_{nk}\mathcal{D}_{nl}\psi_{k} \right.\\
 &\left. +\frac{1}{N_{\mathrm{s}}}\sum\limits_{i=1}^{N_{\mathrm{s}}}\left(\theta_{n}\mathcal{D}_{il}-\mathcal{D}_{il}\mathcal{D}_{nk}\psi_{k}\right) \right. \\
 &\left. +\frac{1}{N_{\mathrm{s}}}\sum\limits_{i=1}^{N_{\mathrm{s}}}\left(\theta_{i}\mathcal{D}_{nl}-\mathcal{D}_{nl}\mathcal{D}_{ik}\psi_{k}\right) \right. \\
 &\left. -\frac{1}{N_{\mathrm{s}}^{2}}\sum\limits_{i,j=1}^{N_{\mathrm{s}}}\left(\theta_{j}\mathcal{D}_{il}-\mathcal{D}_{jk}\mathcal{D}_{il}\psi_{k}\right)\right] \\
\end{split}
 \end{equation}
after sorting terms with and without $\psi_{k}$-term, we find the linear system of equations
\begin{equation}
\begin{split}
 \mathcal{B}^{\mathrm{m}}_{lk}= \sum\limits_{n=1}^{N_{\mathrm{s}}}\frac{2}{\sigma^{2}_{s}} \Biggl[\mathcal{D}_{nl}\mathcal{D}_{nk}&-\frac{1}{N_{\mathrm{s}}}\sum\limits_{i=1}^{N_{\mathrm{s}}}\left(\mathcal{D}_{nk}\mathcal{D}_{il}+\mathcal{D}_{nl}\mathcal{D}_{ik}\right)\\
 &+\frac{1}{N_{\mathrm{s}}^{2}}\sum\limits_{i,j=1}^{N_{\mathrm{s}}}\mathcal{D}_{il}\mathcal{D}_{jk}\Biggl]
\end{split}
 \end{equation}
and data vector
\begin{equation}
\begin{split}
 \mathcal{V}^{\mathrm{m}}_{l}=\sum\limits_{n=1}^{N_{\mathrm{s}}}\frac{2}{\sigma^{2}_{s}} \Biggl[\theta_{n}\mathcal{D}_{nl}&-\frac{1}{N_{\mathrm{s}}}\sum\limits_{i=1}^{N_{\mathrm{s}}}\left(\theta_{n}\mathcal{D}_{il}+\theta_{i}\mathcal{D}_{nl}\right)\\
 &+\frac{1}{N_{\mathrm{s}}^{2}}\sum\limits_{i,j=1}^{N_{\mathrm{s}}}\theta_{j}\mathcal{D}_{il}\Biggl]
 \end{split}
\end{equation}
where $\theta$ is one component of the observed lens-plane position of the multiple images of the system. In order to obtain the 
full multiple image term contribution, one has to sum the contributions from both coordinate components by substituting $\theta$ with $\theta_{1}$ or $\theta_{2}$
and $\mathcal{D}$ with $\mathcal{D}^{1}$ or $\mathcal{D}^{2}$, respectively. If there are more than one multiple-image systems in the lens,
each system contributes a $\chi^{2}$-term.

    In order to obtain the full linear system of equations, all individual contributions to an 
    entry in the coefficient matrix $\mathcal{B}_{lk}$ or in the data vector $\mathcal{V}_{l}$ need to be summed
    \begin{equation}
     \mathcal{B}_{lk}=\mathcal{B}^{\mathrm{w}}_{lk}+\mathcal{B}^{\mathrm{s}}_{lk}+\mathcal{B}^{\mathrm{m}}_{lk}+\mathcal{B}^{\mathrm{s\_reg}}_{lk}+\mathcal{B}^{\mathrm{c\_reg}}_{lk}
    \end{equation}
        \begin{equation}
     \mathcal{V}_{l}=\mathcal{V}^{\mathrm{w}}_{l}+\mathcal{V}^{\mathrm{s}}_{l}+\mathcal{V}^{\mathrm{m}}_{l}+\mathcal{V}^{\mathrm{s\_reg}}_{l}+\mathcal{V}^{\mathrm{c\_reg}}_{l}.
    \end{equation}

  \end{document}